\documentclass[aps,prb,amsmath,amssymb,reprint,superscriptaddress,preprintnumbers,showpacs,intlimits]{revtex4-1}
\usepackage{bm,mathrsfs,txfonts}
\usepackage[mathcal]{euscript}
\usepackage[breaklinks=true,unicode=true,urlcolor = blue,colorlinks = true,citecolor = blue,linkcolor = blue]{hyperref}
\usepackage{graphicx}
\usepackage[subrefformat=parens]{subcaption}
\usepackage{color}

\usepackage{multirow}
\renewcommand{\vec}[1]{\bm{#1}}
\allowdisplaybreaks
\graphicspath{{./}}
\usepackage{pgf}
\usepackage{tikz}
\usetikzlibrary{arrows,shapes,backgrounds}
\usetikzlibrary{positioning,arrows}
\usetikzlibrary{calc}

\newcommand{\makeFilledTikzBoxUp}[5]{%
\coordinate (pointer) at #1;
\coordinate (leftCorner) at #2;
\filldraw[very thin,fill=green!05!white,opacity=0.5,rounded
corners=7pt,xshift=10,yshift=0] %
(leftCorner) -- ($(leftCorner) + (#3/2-0.1,0)$) -- (pointer) --
($(leftCorner) + (#3/2+0.1,0)$) -- ($(leftCorner) + (#3,0)$) --
($(leftCorner) + (#3,-#4)$) --($(leftCorner) + (0,-#4)$) -- cycle;
\node[] at ($(leftCorner) + (#3/2-0.01,-#4/2)$)%
{#5};}

\newcommand{\makeFilledTikzBoxLeft}[5]{%
\coordinate (pointer) at #1;
\coordinate (leftCorner) at #2;
\filldraw[thin,fill=green!05!white,opacity=0.5,rounded
corners=7pt,xshift=10,yshift=0] %
(leftCorner) -- ($(leftCorner) + (#3,0)$) -- ($(leftCorner) + (#3,-#4)$) --
($(leftCorner) + (0,-#4)$) -- ($(leftCorner) + (0,-#4/2-0.1)$) --
(pointer) -- ($(leftCorner) + (0,-#4/2+0.1)$) -- cycle;
\node[] at ($(leftCorner) + (#3/2-0.01,-#4/2)$)%
{#5};}

\newcommand{\makeFilledTikzBoxDown}[5]{%
\coordinate (pointer) at #1;
\coordinate (leftCorner) at #2;
\filldraw[thin,fill=green!05!white,opacity=0.5,rounded
corners=7pt,xshift=10,yshift=0] %
(leftCorner) --  ($(leftCorner) + (#3/2-0.1,0)$) -- (pointer) --
($(leftCorner) + (#3/2+0.1,0)$) --
($(leftCorner) + (#3,0)$) -- ($(leftCorner) + (#3,#4)$) --
($(leftCorner) + (0,#4)$) -- cycle;
\node[] at ($(leftCorner) + (#3/2-0.01,#4/2)$)%
{#5};}

\newcommand{\makeFilledTikzDoubleBoxDown}[6]{%
\coordinate (pointer1) at #1;
\coordinate (pointer2) at #2;
\coordinate (leftCorner) at #3;
\filldraw[thin,fill=green!05!white,opacity=0.5,rounded
corners=7pt,xshift=10,yshift=0] %
(leftCorner) -- ($(leftCorner) + (#4/3-0.1,0)$) -- (pointer1) --
($(leftCorner) + (#4/3+0.1,0)$) -- ($(leftCorner) + (#4/3*2-0.1,0)$) --
(pointer2) --($(leftCorner) + (#4/3*2+0.1,0)$) --
($(leftCorner) + (#4,0)$) --
($(leftCorner) + (#4,#5)$) --($(leftCorner) + (0,#5)$) -- cycle;
\node[] at ($(leftCorner) + (#4/2-0.01,#5/2)$)%
{#6};}

\definecolor{green1}{HTML}{00AA00}

\begin{document}

\title{Regular and chaotic vortex core reversal by a resonant perpendicular
magnetic field}

\author{Oleksandr V. Pylypovskyi}
\email[Corresponding author. Electronic address:]{engraver@univ.net.ua}
\affiliation{Taras Shevchenko National University of Kiev, 01601 Kiev, Ukraine}

\author{Denis D. Sheka}
\email{sheka@univ.net.ua}
\affiliation{Taras Shevchenko National University of Kiev, 01601 Kiev, Ukraine}

\author{Volodymyr P. Kravchuk}
\email{vkravchuk@bitp.kiev.ua}
\affiliation{Institute for Theoretical Physics, 03680 Kiev, Ukraine}

\author{Franz G.~Mertens}
\email{franzgmertens@gmail.com}
\affiliation{Physics Institute, University of Bayreuth, 95440 Bayreuth, Germany}

\author{Yuri Gaididei}
\email{ybg@bitp.kiev.ua}
\affiliation{Institute for Theoretical Physics, 03680 Kiev, Ukraine}
\date{June 17, 2013}

%
%

\begin{abstract}
Under the action of an alternating perpendicular magnetic field the polarity of
the vortex state nanodisk can be efficiently switched. We predict the regular
and \emph{chaotic} dynamics of the vortex polarity and propose a simple
analytical description in terms of a \emph{reduced vortex core} model.
Conditions for the controllable polarity switching are analyzed.
\end{abstract}

\pacs{75.75.-c, 75.78.-n, 75.78.Jp, 75.78.Cd, 05.45.-a}



\maketitle

\section{Introduction}

Investigation of the magnetization dynamics at the nanoscale is a key task of the modern nanomagnetism.\cite{Braun12} One of the typical topologically nontrivial magnetic configurations of a nanoscaled magnet is a magnetic vortex, which can form a ground state configuration of submicron--sized magnetic disk--shaped particles (nanodots). Such a vortex is characterized by a curling divergent--free in--plane configuration with magnetization tangential to the edge surface of the nanoparticle. \cite{Hubert98} The out--of--plane magnetization appears only in a very thin region around the vortex core with about the size of an exchange length (typically about 10 nm for magnetically soft materials \cite{Wachowiak02}). The vortex state is degenerated with respect to the upward or downward magnetization of the vortex core (the vortex polarity $p=\pm1$), hence the vortex polarity can be considered as a bit of information in nonvolatile magnetic vortex random--access memories (VRAM). \cite{Kim08,Yu11a} That is why one
needs to control the vortex polarity switching process in a very fast way.

The vortex polarity switching phenomena were predicted originally for the Heisenberg 2D magnets. \cite{Gaididei99,Gaididei00} The interest to this problem was renewed after an experimental detection of the vortex core reversal in nanodots by an excitation with short bursts of an alternating field, \cite{Waeyenberge06} which opened a possibility to use the vortex state dots as the VRAM. Moreover, this motivated numerous fundamental studies of the vortex core switching mechanism itself.\cite{Braun12}

There are two basic scenarios of the vortex polarity switching. In the first, axially--symmetric (or punch--through) scenario, the vortex polarity is switched due to the direct pumping of axially--symmetric magnon modes. Such a switching occurs, e.g., under the influence of a DC transversal field. \cite{Okuno02,Thiaville03,Kravchuk07a,Vila09} In the second, axially--asymmetric scenario, the switching occurs due to a nonlinear resonance in the system of certain magnon modes with nonlinear coupling. \cite{Kravchuk09,Gaididei10b} Such a scenario is accompanied by the temporary creation and annihilation of vortex--antivortex pairs. \cite{Waeyenberge06} The axially--asymmetric switching occurs, e.g., under the action of different in--plane AC magnetic fields or by a spin polarized current, see Ref.~\onlinecite{Gaididei08b} and references therein.

Recently the interest to the axially--symmetric switching was renewed: using the micromagnetic simulations \citet{Wang12,Yoo12} demonstrated that the vortex polarity reversal can be realized under the action of an alternating perpendicular magnetic field. In this case the resonant pumping of the radial magnon modes initiates the switching at much lower field intensities than by the DC fields.

We have very recently predicted the possibility of the chaotic dynamics of the vortex polarity under the action of the homogeneous transversal AC magnetic field in the 10 GHz range:\cite{Pylypovskyi13b}
\begin{equation} \label{eq:sim:Field}
\vec B(t) = \vec{e}_z B_0 \sin \left(2\pi f t \right).
\end{equation}
In order to describe the switching behavior we proposed in Ref.~\onlinecite{Pylypovskyi13b} the analytical two--parameter cutoff model, which gave us a possibility to describe both deterministic and chaotic behavior of the vortex polarity.

The goal of the current work is to study in detail the vortex dynamics under the action of a perpendicular AC magnetic field: we found a rich vortex polarity dynamical behavior, including the \emph{regular} and \emph{chaotic} regimes of magnetization reversal. In order to analyze the complicated temporal evolution of the vortex polarity we used here the discrete \emph{reduced core model}, \cite{Gaididei99,Gaididei00,Zagorodny03} which allows us to describe different regimes of vortex polarity dynamics, including the resonant behavior, the weakly nonlinear regimes, the reversal dynamics, and the chaotic regime. The reduced core model is another way to treat the discretness effects. As opposed to the cutoff model, the core model is simpler, hence it allows to go further in analytics.

The paper is organized as follows: The full--scale micromagnetic simulations are detailed in Sec.~\ref{sec:sim}. Our diagram of switching events demonstrates regimes of the regular reversal (single, multiple and periodic ones), intermittent and chaotic regimes. In Sec.~\ref{sec:core} we describe the comprehensive vortex core dynamics using a simple collective coordinate model, which provides all features of the full--scale simulations. We propose a way of a unidirectional switching controlled switching in Sec.~\ref{sec:control}. In~Sec.~\ref{sec:conclusion} we state our main conclusions. In Appendix \ref{app:core} we derive the reduced core mode. We use the method of multiple scales to perform a weakly nonlinear analysis of the analytical model in Appendix \ref{app:wna}.

\section{Micromagnetic Simulations of Regular and Chaotic Dynamics}
\label{sec:sim}

\begin{figure*}
\begin{center}
\begin{tikzpicture}[scale=1.5]
\node[right] at (0,2.5) {\includegraphics[width=\textwidth]{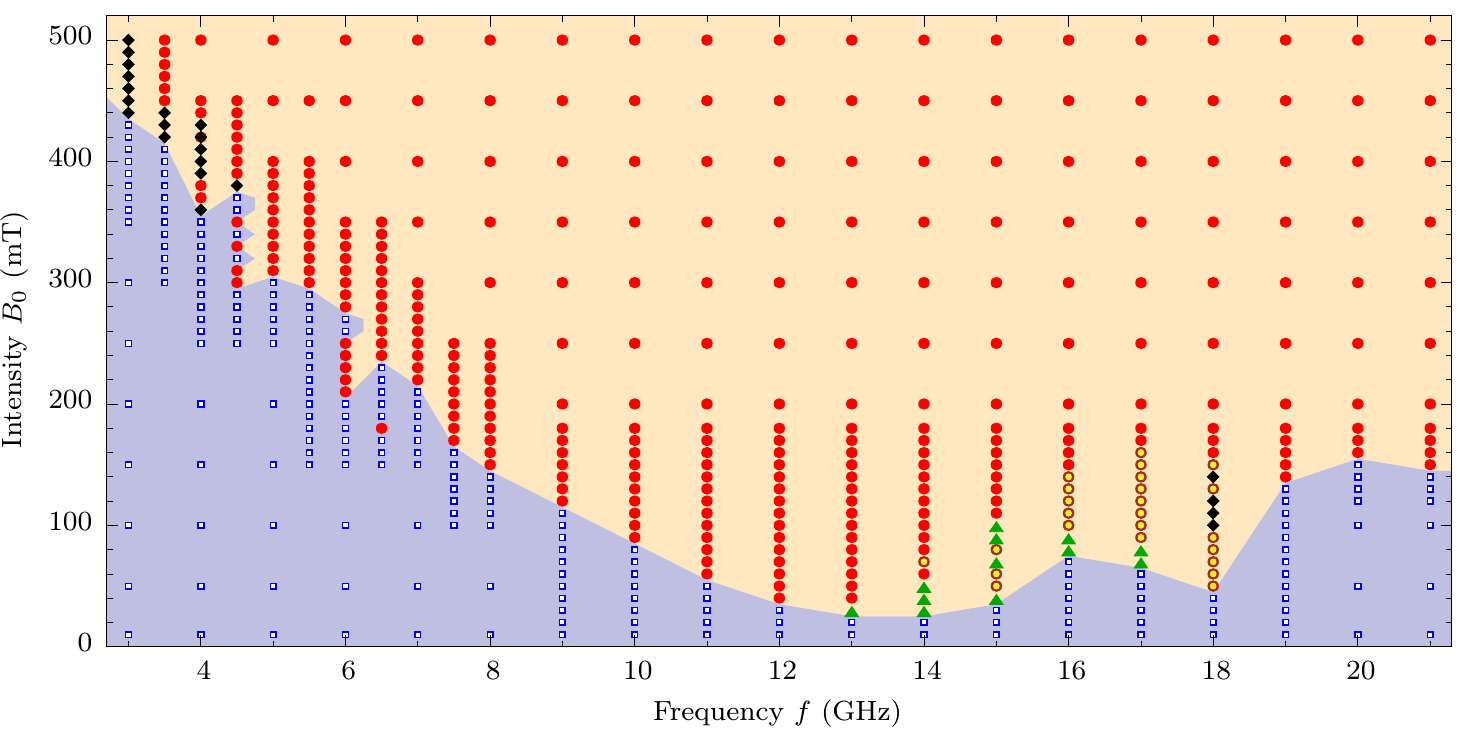}};

\makeFilledTikzBoxDown{(8.7,0.92)}{(0,5.5)}{2.925}{1.9}{%
\includegraphics{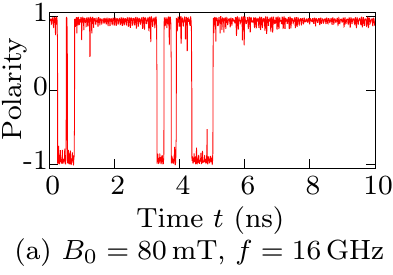}};

\makeFilledTikzDoubleBoxDown{(6.35,0.75)}{(7.5,0.75)}{(3.025,4.475)}{2.925}{
2.925 }{%
\includegraphics{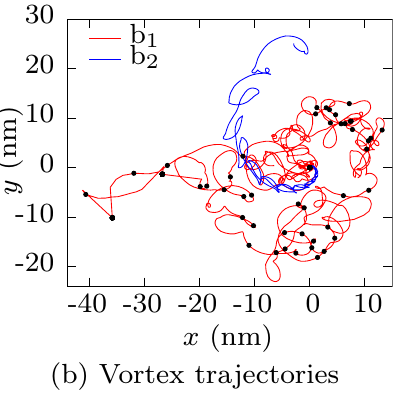}};

\makeFilledTikzBoxDown{(8.7,1.2)}{(6.05,5.5)}{2.925}{1.9}{%
\includegraphics{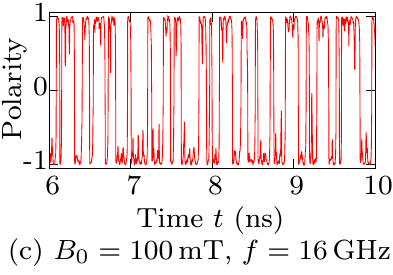}};

\makeFilledTikzBoxDown{(9.96,1.2)}{(9.075,5.5)}{2.925}{1.9}{%
\includegraphics{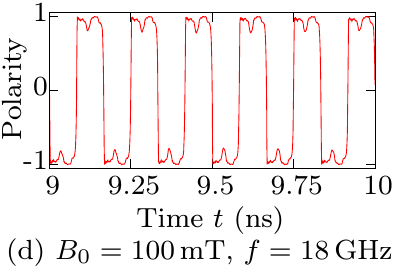}};

\draw[ultra thick,->] (1.7,3)  node[left] {$f/3$} -- (2.,3.4);
\draw[ultra thick,->] (2.4,2.1)  node[left] {$f/2$} -- (2.8,2.3);

\end{tikzpicture}
\end{center}
\caption{(Color online) Switching diagram: open boxes $\textcolor{blue}{\square}$ describe the vortex dynamics without switching and other correspond to parameters where the polarity reversal is observed (the red circles {\Large $\textcolor{red}{\bullet}$} indicate parameters where the vortices escape from the origin during the first 10\,ns, the open circles {\LARGE $\textcolor{yellow}{\bullet}$\,\kern -0.81em\,$\textcolor{brown}{\circ}$} represent parameters where the autocorrelation function \eqref{eq:autocorrelation} rapidly decays, for the diamonds {\large $\blacklozenge$} the switching process is periodical, and the green triangles {\large $\textcolor[rgb]{0.00,0.50,0.00}{\blacktriangle}$} correspond to an intermittent process).
(a) The example of the intermittent process.
(b) The examples of the vortex trajectories in case of the complex vortex 
dynamics. Black points mark the places of the polarity switching. The 
trajectories b$_1$ and b$_2$ correspond to 12\,GHz and 14\,GHz respectively.
(c) The example of the chaotic process.
(d) The example of the regular process.
}
\label{fig:sim:homega}
\end{figure*}

Nowadays the micromagnetic simulations are the inherent tools for the nanomagnetic research.\cite{Kim10} Namely using the numerical simulations it was shown in Refs.~\onlinecite{Wang12,Yoo12} that the resonant perpendicular field forces the vortex core to reverse. Here we perform full--scale micromagnetic simulations to study the complicated vortex core dynamics in details. We consider a disk--shaped nanoparticle (198 nm in diameter and 21 nm in thickness) under the action of the vertical oscillating field \eqref{eq:sim:Field} using an \textsc{OOMMF}\cite{OOMMF} micromagnetic simulator with integration method RK5(4)7FC. The material parameters correspond to Permalloy ($\mathrm{Ni}_{81}\mathrm{Fe}_{19}$) with exchange constant $A = 13$\,pJ, saturation magnetization $M_s = 860$\,kA/m, zero anisotropy coefficient and the Gilbert damping coefficient $\alpha = 0.01$. The mesh cell was chosen to be $3\times 3\times 21$\,nm (the three--dimensional mesh will be discussed at the end of this section).  For all \textsc{
OOMMF} simulations we use as initial state the relaxed vortex with the polarity directed upward and counter--clockwise in--plane magnetization direction.

We also simulated the dynamics of the vortices for the samples with other geometrical parameters: as we expected the qualitative behavior of the system remains the same. \footnote{We also check the polarity dynamics for the samples with the same
diameter and height 33\,nm, with diameter 360\,nm and heights 21\,nm and 33\,nm
under the action of five different pairs of field intensity and frequency:
(60\,mT, 10\,GHz), (60\,mT, 13\,GHz), (80\,mT, 13\,GHz), (100\,mT, 13\,GHz),
(100\,mT, 18\,GHz). The qualitative behavior of the system described in this
section remains the same. Nevertheless the geometrical parameters influence the frequencies of magnon modes and the threshold value of the switching field. Therefore, we expect that the change of the nanodot size shifts the characteristic frequencies of the minimal threshold fields and the threshold field amplitudes, while the qualitative behavior of the system remains the same.}

First of all we examined the eigenfrequencies of the lower axially--symmetric 
spin waves by applying a rectangular pulse with the strength of 30\,mT during 
100\,ps perpendicular to the nanodisk in the vortex state in the same way as in 
Ref.~\onlinecite{Wang12}. Under the action of such a pulse, the 
magnetization  starts to oscillate: a set 
of symmetrical magnon modes $f_{m=0}^{n}$ is excited. Using the fast Fourier 
transformation (FFT) of the $z$--component of the total magnetization, 
typically during $t\in[100\text{\,ps}; 20\text{\,ns}]$, we identified the 
eigenfrequency of the lowest symmetrical mode $f_{m=0}^{n=1} = 13.98$~GHz. This 
value defines the lowest threshold for the polarity switching~\cite{Yoo12}. The 
next nearest peaks in the FFT spectrum correspond to 16.75~GHz and 27.93~GHz.

It is already known\cite{Yoo12} that the vortex polarity switching under the action of the AC field \eqref{eq:sim:Field} occurs in a wide range of field parameters (the field intensities $B_0$ and field frequencies $f$). The minimal field intensity is reached at about the resonance frequency $f_0^1$. In the current study we are interested in the long--time vortex dynamics, which is accompanied by the axially--symmetric polarity reversal mechanism. In all numerical experiments we calculated the polarity and the position of the vortex as functions of time: The vortex position $\vec{R}(t)$ is determined as cross--section of isosurfaces $M_x(\vec{R})=0$ and $M_y(\vec{R})=0$,\cite{Hertel07} and the vortex polarity $p(t)$ is determined as the average $z$--magnetization of four neighbor cells to $\vec{R}(t)$.

To study in details the temporal evolution of the vortex polarity, we simulated the long--time system dynamics with the time step of 1\,ps for a wide range of the field parameters (the field intensity $B_0$ varies from $10$ to $500$\,mT, and the field frequency $f$ changes from $3$ to $21$ GHz).\footnote{The field and frequency increments are varied depending on the position on the diagram of switching events: 10\,mT and 1\,GHz in the range $9\div19$\,GHz and $10\div180$\,mT, 10\,mT and 0.5\,GHz in the range $3\div8.5$\,GHz and $100\div500$\,mT along the border of the diagram, 50\,mT and 1\,GHz for other ranges. For the frequencies $f > 8.5$\,GHz the full time of simulations was 10\,ns [extended till 30\,ns for some particular points $(f,B_0)$] and for frequencies $f \le 8.5$\,GHz the full time of simulations was 5\,ns.}

The results can be summarized in the diagram of dynamical regimes, see Fig.~\ref{fig:sim:homega}. Depending on the field parameters ($B_0$, $f$), one can separate several different dynamical regimes: (i) the absence of the vortex polarity switching, (ii) the chaotic polarity oscillations, (iii) the regular switchings with frequencies depending on the field frequency, (iv) the intermittent switchings, and (v) the complex vortex--magnon dynamics, where the vortex escapes from the origin.

\begin{figure}
\begin{center}
\includegraphics[width=\columnwidth]{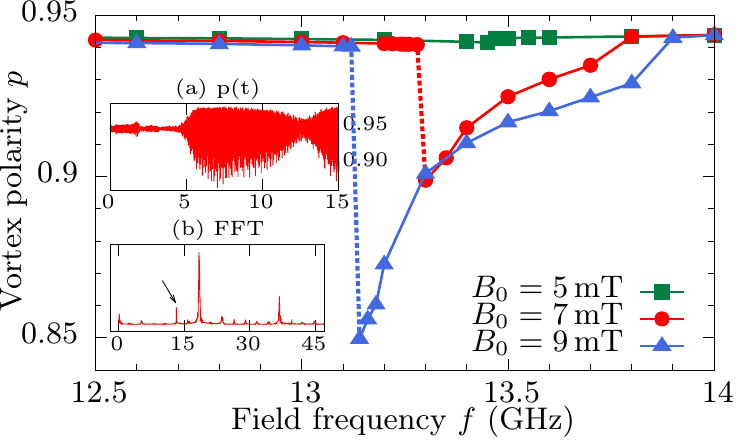}
\end{center}
\caption{(Color online) Nonlinear resonance curves from micromagnetic simulations. Insets: (a) The temporal evolution of the polarity, (b) the FFT spectrum of the vortex polarity for $B_0=9$\,mT, $f=13.2$\,GHz during 15\,ns. Arrow indicates the pumping frequency. }
\label{fig:sim:nonlinAmp}
\end{figure}

(i) We start from a weak field: the field intensity is not strong enough to 
switch the vortex polarity; this regime corresponds to the linear or weakly 
nonlinear oscillations of the vortex polarity (marked as open boxes in 
Fig.~\ref{fig:sim:homega}). The weak pumping of the system (field intensities 
$B_0 \lesssim 5$\,mT, see Fig.~\ref{fig:sim:nonlinAmp}) causes the resonance at 
the frequency $f_0^1$. The increase of the field intensity leads to the 
nonlinear dynamical regime. However, if the field intensity is not strong 
enough, one has a weakly nonlinear regime, which corresponds to the nonlinear 
resonance. Apart from the nonlinear resonance behavior, the strong pumping 
causes the vortex polarity instability,\footnote{Note that the vortex polarity 
instability which correspond to the nonlinear resonance was predicted in our 
previous paper, see Ref.~\onlinecite{Pylypovskyi13b}.} it also causes the shift 
of the main peak in the FFT spectrum (see Fig.~\ref{fig:sim:nonlinAmp}b) and the 
beats in the polarity
oscillations (see Fig.~\ref{fig:sim:nonlinAmp}a).

Let us consider the case when the magnetization reversal occurs. The switching diagram (see Fig.~\ref{fig:sim:homega}) has two well--defined minima. The first one corresponds to the resonant excitation of the radially symmetrical mode $f_0^1$. The second minimum near 18\,GHz, probably, corresponds to the dynamics near the higher resonances\cite{Yoo12}.

(ii) The open circles {\LARGE $\textcolor{yellow}{\bullet}$\,\kern -0.81em\,$\textcolor{brown}{\circ}$} on the switching diagram (see Fig.~\ref{fig:sim:homega}) correspond to the chaotic polarity reversal process. The typical temporal evolution is shown in Fig.~\ref{fig:sim:homega}c. To draw a conclusion about chaotic behavior of the vortex polarity, or more accurately, to make quantitative measures of chaotic dynamics, we use two standard ways: the autocorrelation function for the temporal evolution of the vortex polarity and the Fourier distribution of its frequency spectra.\cite{Moon04}

\begin{figure}
\begin{center}
\includegraphics[width=\columnwidth]{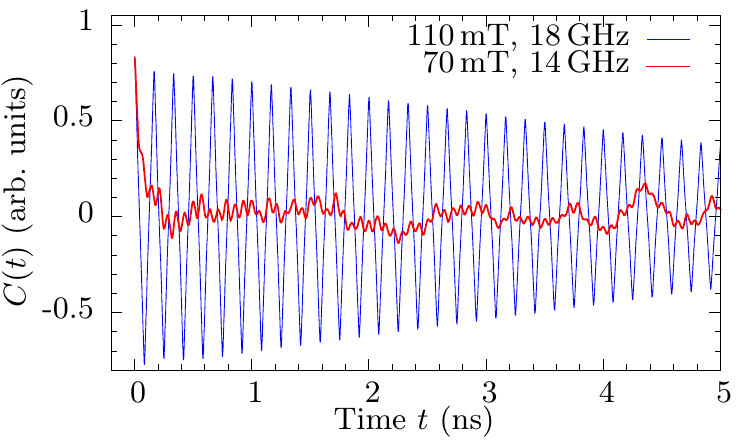}
\end{center}
\caption{(Color online) Autocorrelation function \eqref{eq:autocorrelation} for $B_0=110$\,mT and $f=18$\,GHz (blue dashed line) and  $B_0=70$\,mT and $f=14$\,GHz (red solid line). The first process demonstrates a periodical process with periodical $C(t)$ and the second one demonstrates a chaotic behaviour with rapidly decaying $C(t)$.
}
\label{fig:sim:autocor}
\end{figure}

First, we define the autocorrelation function of the vortex polarity signal \begin{equation} \label{eq:autocorrelation} C(t_i) = \frac{1}{N}\sum_{j=1}^N p(t_{i+j}) p(t_j),\quad i=\overline{1,N} \end{equation} for the discretized time $t_j = j t_0$ with the step $t_0=1$\,ps, with the boundary values assumed as zero. It is well known\cite{Moon04} from the correlational analysis, when a signal is chaotic, information about its past origins is lost, i.e. the signal is only correlated with its recent part: the autocorrelation function decays very rapidly, $C(t)\to0$ as $t\to\infty$.\cite{Moon04} For a periodical signal, the autocorrelation function is a periodic too. A typical example is presented in Fig.~\ref{fig:sim:autocor}: the autocorrelation function $C(t)$ is aperiodic and sharply decays for the applied magnetic field 70\,mT with the frequency 14\,GHz, which corresponds to the chaotic dynamics. The autocorrelation for the regular dynamics demonstrates the oscillations under the action of $B_0=110$\,mT with
$f=18$\,GHz.

\begin{figure}
\begin{center}
\includegraphics[width=\columnwidth]{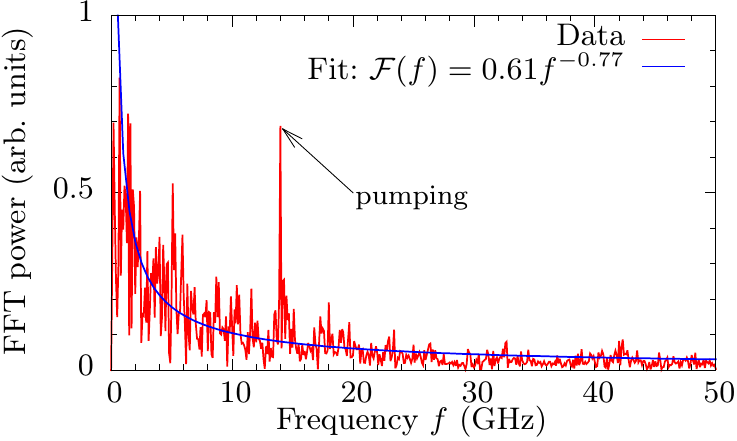}
\end{center}
\caption{(Color online) FFT spectrum of the vortex polarity ($B_0=70$\,mT, $f=14$\,GHz): solid line corresponds to the numerical data, the dashed line is the fitting to the pink noise. }
\label{fig:sim:fourier}
\end{figure}

The second way is to calculate the Fourier spectrum of a chaotic signal. A typical FFT signal is presented in the Fig.~\ref{fig:sim:fourier}. It is distinctive for the chaotic regime that the continuous spectrum dominates the discrete spikes (one can identify in the Fig.~\ref{fig:sim:fourier} only one discrete spike at the pumping frequency). The fitting of such a signal demonstrates typical pink noise behavior with a power law decay of the spectrum, $\mathcal{F}(f) \propto 1/f^\beta$ with $\beta\approx 0.77$.\footnote{As opposed frequency--independent white noise, the pink noise is characterized by the power law decay with $1/f^\beta$ spectrum, where $\beta\in(0,2)$, see Ref.~\onlinecite[p.~102]{Chen04}.}

(iii) The regular oscillations of the vortex polarity appear in the high frequency regime, see the black diamonds on the switching diagram (see Figs.~\ref{fig:sim:homega},~\ref{fig:sim:homega}d). We have detected the periodical motion of the vortex polarity using the pumping frequency 18\,GHz with the field intensities higher than 100\,mT. The main peak in the FFT spectrum corresponds to 6\,GHz, \emph{i.~e.} it occurs at $f/3$ of the pumping. Other spikes with decaying intensities appear with steps of 6\,GHz. We compare the autocorrelation functions for the regular and chaotic oscillations, see Fig.~\ref{fig:sim:autocor}. In contrast to the chaotic regime, $C(t)$ for periodic oscillations exhibits a high periodicity with a slowly decaying amplitude due to the finite observation time.

\begin{figure}
\begin{center}
\includegraphics[width=\columnwidth]{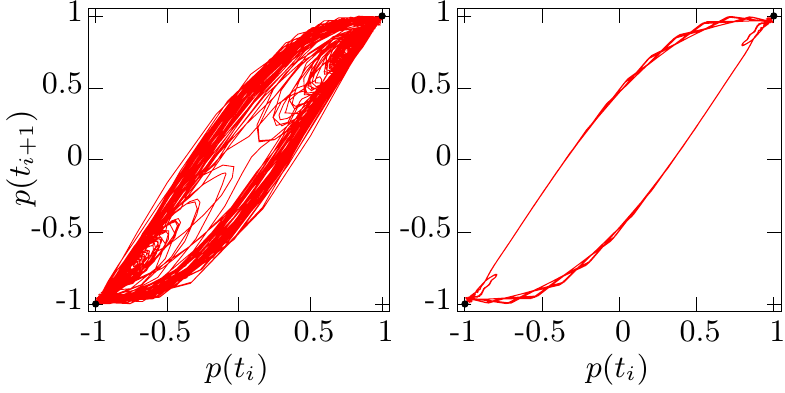}\\
(a) $B_0 = 100$\,mT, $f=16$\,GHz \hfill (b) $B_0 = 100$\,mT, $f=18$\,GHz
\end{center}
\caption{(Color online) (a) The pseudo--phase diagram for the chaotic vortex polarity dynamics (the time step 1\,ns, the filled circles mark polarities $p=\pm 1$); (b) the same diagram for the regular dynamics.}
\label{fig:sim:pseudofield}
\end{figure}

In order to compare the temporal dynamics of the polarity in chaotic and regular regimes, we calculate the pseudo--phase trajectories. The method of the pseudo--phase space is usually used when only one variable [the discretized vortex polarity $p(t_i)$ in our case] is measured:\cite{Moon04} the pseudo--phase--space plot can be made using $p(t_i)$ and its future value $p_{t_{i+1}}$, where the absolute value of the time step $t_{i+1}-t_i$ affects only the shape of the trajectory. In the case of chaotic dynamics, one has open trajectories in pseudo--phase--space $\bigl( p(t_i), p(t_{i+1}) \bigr)$, see the Fig.~\ref{fig:sim:pseudofield}(a). In the regular case, pseudo--phase trajectories are closed, see the  Fig.~\ref{fig:sim:pseudofield}(b). Both trajectories are shown for the first 10\,ns of the dynamics: in the first case the trajectory every time makes a new loop in a different place and in the second case all loops coincide. Below in Sec.~\ref{sec:core} we construct the phase trajectories for the
theoretical model of our system (see Figs.~\ref{fig:core:homega}a, \ref{fig:core:homega}d).

(iv) The green triangles on the switching diagram (see Fig.~\ref{fig:sim:homega}) correspond to an intermittent process. The typical example of the temporal dynamics in such a regime is plotted in the Fig.~\ref{fig:sim:homega}a: the vortex state can retain its polarity for a relatively long time of a few nanoseconds; after that multiple reversal processess occur during $50-100$\,ns. Note that in the vicinity of other regimes in the switching diagram we observed that the vortex polarity, after a few switches, can be `frozen' for the rest of the observation time. For example, two switching events occur during the first 1.2\,ns ($B_0 = 30$\,mT and $f=14$\,GHz, see Fig.~\ref{fig:sim:homega}a), after that the dynamical polarity has only weak oscillations. A similar picture occurs for $B_0 = 70$\,mT and $f=17$\,GHz, where after three reversals during the first nanosecond the resulting polarity remains negative. Since the reversals occur only at the beginning, one can conclude that this occurs because the field is
not switched on smoothly.

(v) The last regime corresponds to the field parameters ($B_0$, $f$), where the vortex escapes from the system origin on a long time scale (see the red circles in Fig.~\ref{fig:sim:homega}). Typically, the vortex starts to move during the first 10\,ns. The detailed analysis shows that the switching scenario differs essentially from the above mentioned one: the magnetization reversal is mediated by the transient creation and annihilation of a vortex--antivortex pair \cite{Waeyenberge06} (for details of the axially--asymmetric switching mechanism see Ref.~\onlinecite{Gaididei08b} and references therein). Two examples of the possible trajectories are shown in Fig.~\ref{fig:sim:homega}b: The trajectory b$_1$ corresponds to the chaotic motion, which is accompanied by numerous reversal events. In the regular regime the vortex trajectory has a smooth shape (b$_2$). When the vortex stays in the center of the sample, polarity switching is accompanied by generation of the radially symmetrical modes. After some time of
observation, a new 4-fold symmetry occurs around the vortex, which was mentioned in the Ref.~\onlinecite{Yoo12} and linked with the square mesh symmetry used in the OOMMF. When the vortex moves from the center, the switching scenario is changed: a pair antivortex--new vortex is created and the antivortex annihilates with the old vortex. The further magnon dynamics becomes unpredictable.

We performed very long--time simulations (up to 30\,ns) for all parameters from the switching diagram, where the vortex does not leave the disk center (see Fig.~\ref{fig:sim:homega}): the vortex motion was found for all parameters with $f< 17$\,GHz. For higher frequencies (e.g., for $f=18$\,GHz and $B_0=100$\,mT) the small oscillations of the vortex position were observed only for $t>29$\,ns. In the prolonged simulations (iv) the vortex polarity does not change its value during the time of observation in agreement with the conclusion made above that the field is sharply switched on.

\label{page:features}
The switching diagram for the low--frequency range has several new features. One can identify from the plot two local minima (4.5\,GHz and 6\,GHz), which correspond to resonances for fractional frequencies ($\frac13f_0^1$ and $\frac12f_0^1$). The strong field causes the vortex polarity reversal, which corresponds to the quasi--static regime and the lower fields cause the escape of the vortex from the system origin. We checked the idea about the quasi--static regime by computing the threshold value for the static field, which is in our case 611\,mT, cf. Refs.~\onlinecite{Thiaville03,Kravchuk07a}.

\begin{figure}
\begin{center}
\includegraphics[width=\columnwidth]{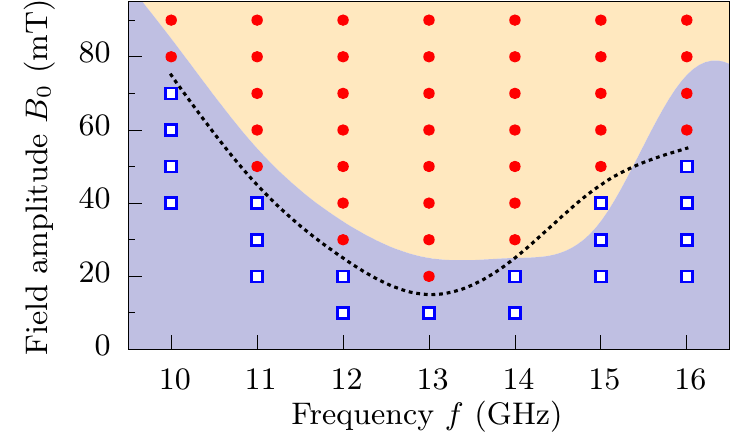}
\end{center}
\caption{(Color online) Comparison of 2D and 3D simulations. Symbols correspond to 3D simulations: The red circles describe the switching process, the open blue boxes describe the dynamics without switching. The dotted line represents the border of the switching region for 3D simulations, colored domains correspond to 2D simulations, see Fig.~\ref{fig:sim:homega}.}
\label{fig:sim:homegaDiagram3D}
\end{figure}

It should be noted that in all simulations discussed above we used the effective 2D mesh $3\times 3\times 21$\,nm. In order to check our assumption about the uniform magnetization distribution along the thickness z--coordinate, we also performed 3D OOMMF simulations with the mesh size $3 \times 3 \times 3$\,nm. One can see that eigenfrequencies and boundaries of dynamical regimes are slightly influenced by the nonhomogeneous distribution along the z--coordinate, see Fig.~\ref{fig:sim:homegaDiagram3D}.

It is known that the vortex reversal under the action of a perpendicular static field is accompanied by the temporal creation and annihilation of a Bloch point: the switching process, as a rule, is mediated by the creation of two Bloch points, however, the single Bloch point scenario was also mentioned. \cite{Thiaville03} The Bloch point propagation during the polarity reversal under the ac perpendicular field was also mentioned by \citet{Yoo12}. It should be noted that the Bloch point is a 3D micromagnetic singularity, hence it does not exist in 2D simulations.

The dynamics of the Bloch point is not well studied to the moment. During the switching process we observed a new 3D picture of the switching, where some switching events are not completed: the vortex near one face surface rapidly reverses its polarity to opposite and returns back, while the vortex near the second face surface saves its polarity during this time.

\section{Description of Different Dynamical Regimes}
\label{sec:core}

To gain some insight to the switching mechanism, we need a model which allows the magnetization reversal process. It is worth reminding that in the continuum limit the vortex states with different polarities are separated by an infinite barrier. In the spin lattice the barrier becomes finite \cite{Wysin94} and the reversal can occur. It is already known from our previous paper\cite{Pylypovskyi13b} that the dominating contribution to the switching mechanism is caused by the exchange interaction inside the vortex core. That is why to describe the polarity reversal process we use here the discrete \emph{reduced core model}, which was initially introduced by \citet{Wysin94} for the vortex instability phenomenon. Later the vortex core model was developed to analyze the vortex polarity switching in Heisenberg magnets.\cite{Gaididei99,Gaididei00,Zagorodny03}

One has to note that the reduced core model does not pretend a quantitative agreement with simulations. It is the simplest model which allows to describe a rich variety of different regimes of vortex polarity dynamics, including the resonant behavior, the weakly nonlinear regimes, the reversal dynamics, and the chaotic regime.

We consider the anisotropic classical Heisenberg disk shaped system with thickness $L_z$ and the radius $L$, assuming that the magnetization of the magnet is uniform along the thickness.
In terms of the normalized magnetic moment
\begin{equation}\label{eq:m}
\vec{m}_{\vec{n}} = \left(\sqrt{1-m_{\vec{n}}^2}\cos\phi_{\vec{n}}, \sqrt{1-m_{\vec{n}}^2} \sin\phi_{\vec{n}}, m_{\vec{n}}\right)
\end{equation}
the energy of such a magnet with the account of the interaction with magnetic field reads
\begin{equation} \label{eq:Energy}
\begin{split}
E &= - \frac{A L_z}{2} \sum_{(\vec{n},\vec{\delta})} \left[
\vec{m}_{\vec{n}}\cdot \vec{m}_{\vec{n}+\vec{\delta}} - (1-\lambda)
m_{\vec{n}}^z m_{\vec{n}+\vec{\delta}}^z \right]\\
&- a_0^2 M_s L_z \sum_{\vec{n}} \vec{m}_{\vec{n}}\cdot \vec{B}(t),
\end{split}
\end{equation}
where the vector $\vec{\delta}$ connects nearest neighbors of the three--dimensional cubic lattice with the lattice constant $a_0$, $A$ is the exchange constant, the parameter $\lambda\in(0,1)$ is the effective anisotropy constant, and $M_s$ is the saturation magnetization. According to this model the planar vortex is stable when $\lambda < \lambda_c$, where $\lambda_c \approx 0.72$ for the square lattice.\cite{Wysin94}
In a cylindrical frame of reference $(r,\chi,z)$ the planar vortex distribution is described by
\begin{subequations} \label{eq:vortex}
\begin{equation} \label{eq:planar-vortex}
m_v = 0, \qquad \phi_v = \chi + \mathcal{C},
\end{equation}
where $\mathcal{C} = \pm \pi/2$ is a vortex chirality (we use here the positive
sign in calculations below). When $\lambda>\lambda_c$, the nonplanar vortex
appears, which is characterized by the well--localized out--of--plane
magnetization $m_v\neq0$.\cite{Wysin94}

\begin{figure}
\begin{center}
\begin{tikzpicture}[scale=1.5]


\draw[step=1.cm,black,thick] (-0.5,-0.5) grid (3.5,3.5);

\draw[black,dashed,line width=1pt, xshift=17.5mm,yshift=7.5mm] (0,0) --
(0.9, 0.9);
\draw[red,dashed,line width=1pt, xshift=17.5mm,yshift=9.1mm] (0,0) --
(1.20344867377942, 0.4143806095590807);

\draw[fill=red] (1,1) circle (3pt);
\draw[fill=red] (1,2) circle (3pt);
\draw[fill=red] (2,1) circle (3pt);
\draw[fill=red] (2,2) circle (3pt);

\draw[fill=black] (0,0) circle (2pt);
\draw[fill=black] (0,3) circle (2pt);
\draw[fill=black] (3,0) circle (2pt);
\draw[fill=black] (3,3) circle (2pt);

\draw[fill=black] (1,0) circle (2pt);
\draw[fill=black] (2,0) circle (2pt);
\draw[fill=black] (3,1) circle (2pt);
\draw[fill=black] (3,2) circle (2pt);
\draw[fill=black] (2,3) circle (2pt);
\draw[fill=black] (1,3) circle (2pt);
\draw[fill=black] (0,2) circle (2pt);
\draw[fill=black] (0,1) circle (2pt);

\draw[black,line width=2pt, ->, xshift=6.2053mm,yshift=1.2649mm] (0,0) --
(0.75895,-0.25298);
\draw[black,line width=2pt, ->, xshift=16.2053mm,yshift=-1.2649mm] (0,0) --
(0.75895,0.25298);

\draw[black,line width=2pt, ->, xshift=28.7351mm,yshift=6.2053mm] (0,0) --
(0.25298, 0.75895);
\draw[black,line width=2pt, ->, xshift=31.265mm,yshift=16.205mm] (0,0) --
(-0.25298, 0.75895);

\draw[black,line width=2pt, ->, xshift=23.795mm,yshift=28.735mm] (0,0) --
(-0.75895,0.25298);
\draw[black,line width=2pt, ->, xshift=13.795mm,yshift=31.265mm] (0,0) --
(-0.75895,-0.25298);

\draw[black,line width=2pt, ->, xshift=-1.2649mm,yshift=13.7947mm] (0,0) --
(0.25298, -0.75895);
\draw[black,line width=2pt, ->, xshift=1.2649mm,yshift=23.7947mm] (0,0) --
(-0.25298, -0.75895);

\draw[black,line width=2pt, ->, xshift=27.5mm,yshift=-2.5mm] (0,0) -- (0.56569,
0.56569);
\draw[black,line width=2pt, ->, xshift=32.5mm,yshift=27.5mm] (0,0) -- (-0.56569,
0.56569);
\draw[black,line width=2pt, ->, xshift=2.5mm,yshift=32.5mm] (0,0) -- (-0.56569,
-0.56569);
\draw[black,line width=2pt, ->, xshift=-2.5mm,yshift=2.5mm] (0,0) -- (0.56569,
-0.56569);

\draw[red,line width=3pt, ->, xshift=17.5mm,yshift=9.2mm] (0,0) --
(0.7564148604794535, 0.26045452356572535);
\draw[red,line width=3pt, ->, xshift=20.8mm,yshift=17.5mm] (0,0) --
(-0.26045452356572535, 0.7564148604794535);
\draw[red,line width=3pt, ->, xshift=12.5mm,yshift=20.8mm] (0,0) --
(-0.7564148604794536, -0.2604545235657251);
\draw[red,line width=3pt, ->, xshift=9.2mm,yshift=12.5mm] (0,0) --
(0.26045452356572507, -0.7564148604794536);

\draw[red,densely dashed,line width=0.75pt, <->, xshift=27mm,yshift=12.5mm]
(0,0) arc (20:42:8mm);

\node[xshift=42mm,yshift=22mm,color=red] at (0,0) {$\psi$};

(0,0) arc (100:170:3mm);
(0,0) arc (190:260:3mm);
(0,0) arc (280:350:3mm);
(0,0) arc (10:80:3mm);

\node[xshift=12mm,yshift=12mm] at (0,0) {4};
\node[xshift=33mm,yshift=12mm] at (0,0) {1};
\node[xshift=33mm,yshift=33mm] at (0,0) {2};
\node[xshift=12mm,yshift=33mm] at (0,0) {3};

\node[xshift=12mm,yshift=-3mm] at (0,0) {12};
\node[xshift=33mm,yshift=-3mm] at (0,0) {5};
\node[xshift=48mm,yshift=12mm] at (0,0) {6};
\node[xshift=48mm,yshift=33mm] at (0,0) {7};
\node[xshift=33mm,yshift=48mm] at (0,0) {8};
\node[xshift=12mm,yshift=48mm] at (0,0) {9};
\node[xshift=-3mm,yshift=33mm] at (0,0) {10};
\node[xshift=-3mm,yshift=12mm] at (0,0) {11};
\end{tikzpicture}
\end{center}
\caption{(Color online) Schematic of the reduced core model: Thick red arrows (numbers $\overline{1,4}$) indicate free magnetic moments and thin black ones (numbers $\overline{5,12}$) indicate fixed magnetic moments with $m_{\vec n}^z = 0$. The turning phase $\psi$ describes the deviation of the magnetization angle from the equilibrium, see Appendix~\ref{app:core} for details.}
\label{fig:core-schematic}
\end{figure}
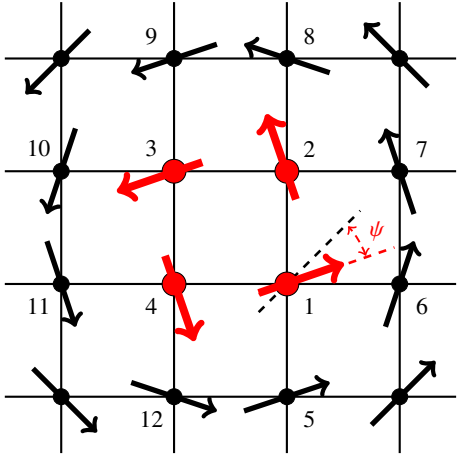

In the reduced core approach, we suppose that only the four magnetic moments of the first coordinate shell can vary, forming the vortex core; all the other moments are fixed in the sample plane in a vortex configuration \eqref{eq:planar-vortex}, see Fig.~\ref{fig:core-schematic}. By symmetry, all four moments are characterized by the same out--of--plane magnetization $\mu$ and equal in--plane phase $\psi$, which is determined as a deviation from the vortex configuration. Therefore, the magnetization distribution of the first coordinate shell is describe as follows:
\begin{equation} \label{eq:magnMomTurnPh}
m_{i}^z = \mu,\quad \phi_{i} =  \chi + \mathcal{C} + \psi, \quad i=\overline{1,4}.
\end{equation}
\end{subequations}
We consider the core magnetization $\mu$, which has the meaning of the \emph{dynamical vortex polarity}, and the in--plane \emph{turning phase} $\psi$ as two collective variables.

The energy of the model, normalized by $\epsilon=8AL_z\lambda$ has the form (see Appendix \ref{app:core} for details):
\begin{equation} \label{eq:E-core}
\mathscr{E} = - \frac{\mu^2}{2} - \Lambda\sqrt{1-\mu^2}\cos\psi - \mu h
\sin\omega \tau,
\end{equation}
where we introduced the reduced anisotropy parameter $\Lambda = 2/(\lambda\sqrt{5})$, the reduced field intensity $h = a_0^2M_sB_0/(2A\lambda)$, the reduced field frequency $\omega = 2\pi f M_s/(\epsilon\gamma)$, and the rescaled time $\tau = \epsilon\gamma t/M_s$. We use $\Lambda = 0.9415$ ($\lambda=0.95$) and $\eta = 0.002$ in a majority of numerical calculations below. Note that such a value of the $\Lambda$--parameter is chosen for illustrative purposes, it does not fit to the correct material parameters from simulations.

The magnetization dynamics in the reduced core model can be described by the following equations (see Appendix \ref{app:core} for details):
\begin{equation} \label{eq:core:motion}
\begin{split}
\dot \mu  &= \Lambda\sqrt{1-\mu^2}\sin\psi + \eta \Bigl[ \mu(1-\mu^2)\\
          & - \Lambda \mu\sqrt{1-\mu^2} \cos\psi + h(1-\mu^2) \sin \omega\tau
\Bigr],\\
\dot \psi &=  \mu - \frac{\Lambda \mu \cos\psi}{\sqrt{1-\mu^2}} + h\sin
\omega\tau - \frac{\eta \Lambda \sin\psi}{\sqrt{1-m^2}},
\end{split}
\end{equation}
where the overdot means the derivative with respect to $\tau$ and $\eta$ is a Gilbert damping coefficient. The ground state of the model corresponds to
\begin{equation} \label{eq:ground-state}
\mu_0 = \pm \sqrt{1-\Lambda^2}, \qquad \psi_0 = 0.
\end{equation}
In terms of the core model two opposite values of $\mu_0$ describe vortices
with opposite polarities $\mu_0$.

\begin{figure}
\begin{center}
\includegraphics[width=\columnwidth]{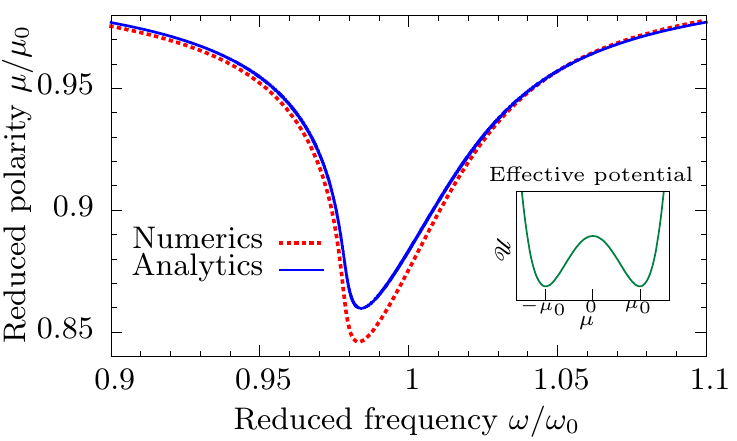}
\end{center}
\caption{(Color online) Nonlinear resonance curve: The numerical solution of Eqs.~\eqref{eq:core:motion} (the dashed line) and the analytical solution \eqref{eq:core:resonanceCurve} (the solid line) for the following parameters: $\Lambda = 0.9415$, $h=0.0002$, $\eta = 0.01$, $\mu(0) = 0.337$, $\psi(0) = 0$. The effective double--well potential $\mathscr{U}$, see \eqref{eq:eff-L}, is plotted in the inset. }
\label{fig:core:nonlinAmp}
\end{figure}
\begin{figure*}
\begin{center}
\begin{tikzpicture}[scale=1.5]
\node[right] at (0,2.5)
{\includegraphics[width=\textwidth]{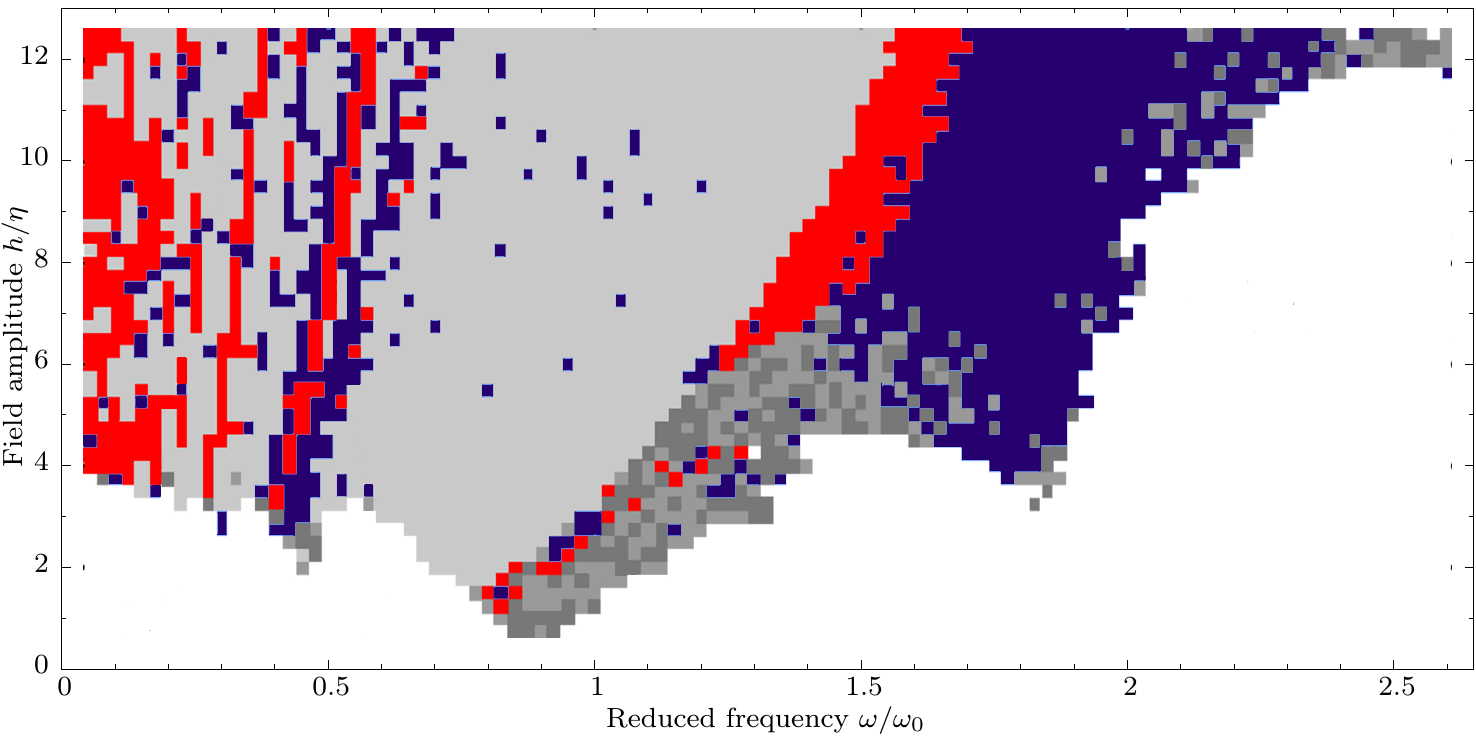}};

\makeFilledTikzBoxDown{(1.9,4.12)}{(0,5.5)}{2.925}{1.9}{%
\includegraphics{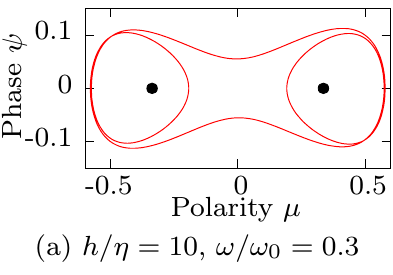}};

\makeFilledTikzBoxDown{(1.85,4.17)}{(3.025,5.5)}{2.925}{1.9}{%
\includegraphics{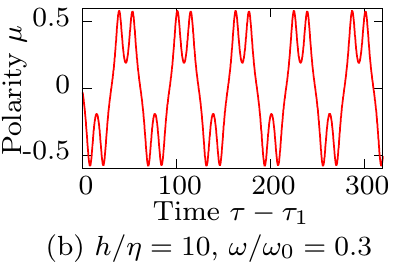}};

\makeFilledTikzBoxDown{(7.05,4.1)}{(6.05,5.5)}{2.925}{1.9}{%
\includegraphics{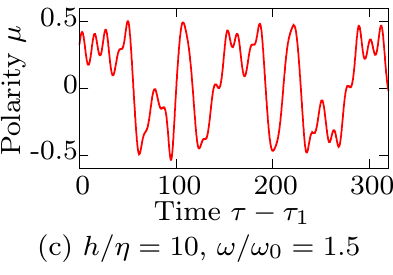}};

\makeFilledTikzBoxDown{(7.,4.15)}{(9.075,5.5)}{2.925}{1.9}{%
\includegraphics{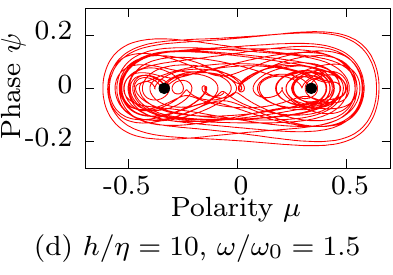}};

\makeFilledTikzBoxUp{(2.77,3.7)}{(0.75,0.8)}{3}{1.9}{%
\includegraphics{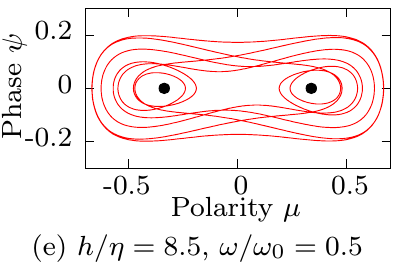}};

\makeFilledTikzBoxLeft{(4.6,0.87)}{(8.9,2.21)}{3}{1.9}{%
\includegraphics{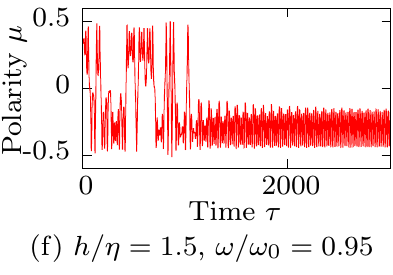}};

\end{tikzpicture}
\end{center}
\caption{(Color online) Map of dynamical regimes in the coordinates 
$\omega/\omega_0-h/\eta$. $\eta = 0.002$, other parameters are the same as in 
Fig.~\ref{fig:core:nonlinAmp}. The red region corresponds to the chaotic 
dynamics. Three scales of grey indicate (in order of growing intensity) the 
switching process with one stable focus on the Poincar\'{e} diagram and the 
final dynamics near the upward or downward polarity. The blue region 
corresponds to the dynamics with more than one stable focus on the Poincar\'{e} 
diagram.
(a) The phase diagram (projection) for the regular dynamics. The black circles indicate the equilibrium polarities $\pm \mu_0$.
(b) Example of the regular dynamics, including the $f/3$ peak in FFT spectrum.
(c) Example of the chaotic oscillations. $\tau_1 = 186\cdot 10^3$.
(d) The same as (a) for the chaotic dynamics.
(e) The same as (b) for the dynamics with 5 stable focuses on the Poincar\'{e} map.
(f) Example of a process with the final state near $-\mu_0$.}
\label{fig:core:homega}
\end{figure*}

Let us start our analysis with a system without damping, $\eta=0$. Supposing that the turning phase is small enough, $|\psi|\ll1$, one can easily exclude $\psi$ from the consideration. In this case the Eqs.~\eqref{eq:core:motion} correspond to the effective Lagrangian
\begin{equation} \label{eq:eff-L}
\begin{split}
\mathscr L = & \frac{\mathscr M \dot\mu^2}{2} - \mathscr{U}(\mu) + \mu h \sin
\omega\tau, \\
\mathscr M = & \frac{1}{\Lambda \sqrt{1-\mu^2}},\qquad \mathscr{U}(\mu) = -
\frac{\mu^2}{2} - \Lambda\sqrt{1-\mu^2}.
\end{split}
\end{equation}
This simplification allows us to interpret the complicated dynamics as the motion of a particle with variable mass $\mathscr M$ in the double--well potential $\mathscr U(\mu)$ under a periodic pumping, see the inset in the Fig.~\ref{fig:core:nonlinAmp}. The linear oscillations near the equilibrium state correspond to the harmonic oscillations of the effective particle in one of the wells; the eigenfrequency of such oscillations is
\begin{equation} \label{omega0}
\omega_0 = \sqrt{1-\Lambda^2}.
\end{equation}

Let us describe the nonlinear regime of the dynamics. In spite of the small damping in the system, its value can be comparable with the pumping intensity. Therefore we consider below the full set of the model equations Eqs.~\eqref{eq:core:motion}. To analyze the weakly nonlinear regime, we use the multiscale perturbation method\cite{Kevorkian81,Nayfeh85,Nayfeh08}. When the field intensity is much less than the frequency detuning ($h \ll |\omega - \omega_0|$), we can limit ourselves to a three--scale expansion in the form
\begin{equation} \label{eq:core:smallAmp}
\begin{split}
\mu = \mu_0 + \sum_{n = 1}^3 \varepsilon^n \mu_n(T_0, T_1, T_2),\quad T_n =
\varepsilon^n\tau,\\
\psi = \sum_{n = 1}^3 \varepsilon^n \psi_n(T_0, T_1, T_2),\quad \omega =
\omega_0 + \omega_\pm,\\
\omega_\pm = \varepsilon^2 \omega_2,\quad \eta = \varepsilon^2 \eta_2, \quad h =
\varepsilon^3h_3.
\end{split}
\end{equation}
Using such the expansion, one can derive from Eqs.~\eqref{eq:core:motion} the resonance curve $\omega_\pm(h)$ as the solution of the following equation
\begin{equation} \label{eq:core:resonanceCurve}
\begin{split}
h^2\Lambda^4 & =a^2\left(2\sqrt{1-\Lambda ^2}\omega_{\pm}+\frac{2+\Lambda
^2}{2\Lambda ^2}a^2\right)^2\\
& +{\eta ^2\left(1-\Lambda ^2\right)\left(2-\Lambda ^2\right)^2}a^2,
\end{split}
\end{equation}
with $|a|$ being the amplitude of oscillations, see Appendix~\ref{app:wna} for details.  The typical nonlinear resonance curve is shown in the Fig.~\ref{fig:core:nonlinAmp}, cf. Fig.~\ref{fig:sim:nonlinAmp}.

We analyze the strongly nonlinear regimes solving numerically Eqs.~\eqref{eq:core:motion} in a wide range of parameters ($\omega,h$), see the diagram of switching events in the Fig.~\ref{fig:core:homega}. The absolute minimum in the diagram corresponds to the switching in the range near the resonance frequency $\omega_0$. Other local minima correspond to resonances at double frequency, $2\omega_0$, and sub--harmonics, $\omega_0/2$ and $\omega_0/3$. Note that all resonance frequencies are shifted in  the low--frequency direction due to the nonlinear resonance.

We classified the dynamical regimes by using the method of Poincar\'{e} maps ($15\cdot 10^3$ points per map). We constructed such maps for each pair ($\omega$, $h$) where the switching takes place. One can separate four oscillation regimes related to the corresponding regimes in the OOMMF simulations (Sec.~\ref{sec:sim}) with vortex dynamics in the center of the sample: (i) the absence of switching, (ii) the chaotic dynamics, (iii) the regular polarity oscillations between two polarities $\pm \mu_0$, (iv) the switching with final oscillations around one of the points $(\pm \mu_0, 0)$ in the coordinates $(\mu,\psi)$.

(i) The border between the switching region and the region, where field or frequency are not enough for jumps between $(\pm \mu_0,0)$, shows a few well-defined resonance minima corresponding to resonances at $\omega_0/3$, $\omega_0/2$, $\omega_0$ and $2\omega_0$.

\begin{figure*}
\begin{center}
\begin{subfigure}{0.32\textwidth}
\includegraphics[width=\columnwidth]{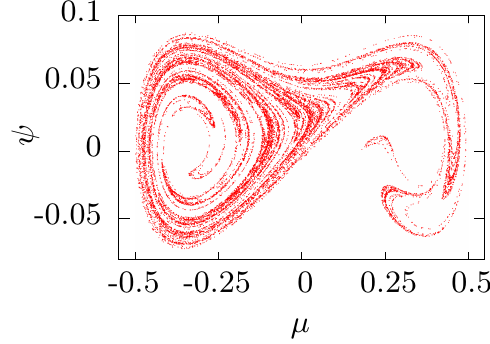}
\caption{$\omega/\omega_0 = 0.05$, $P=2\cdot 10^4$}
\label{fig:poincareEvolution1}%
\end{subfigure}%
\hfill
\begin{subfigure}{0.32\textwidth}
\includegraphics[width=\columnwidth]{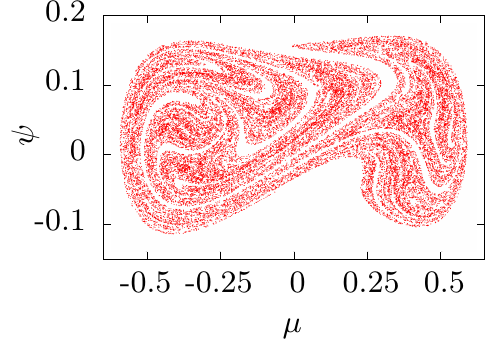}
\caption{$\omega/\omega_0 = 0.35$, $P=3\cdot 10^4$}
\label{fig:poincareEvolution2}%
\end{subfigure}%
\hfill
\begin{subfigure}{0.32\textwidth}
\includegraphics[width=\columnwidth]{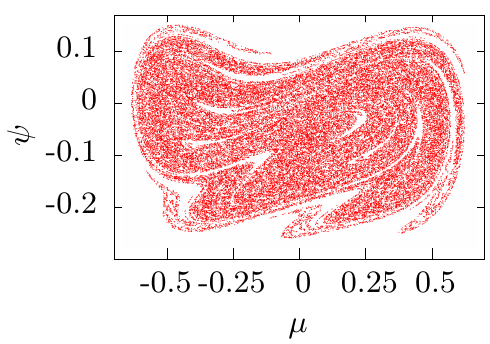}
\caption{$\omega/\omega_0 = 1.6$, $P=5\cdot 10^4$}
\label{fig:poincareEvolution3}%
\end{subfigure}%
\end{center}
\caption{(Color online) Evolution of strange attractors with change of pumping frequency. $h/\eta = 10.5$, other parameters are the same as in Fig.~\ref{fig:core:homega}. The number $P$ means number of points on the corresponding Poincar\'{e} map.}
\label{fig:core:poincareEvolution}
\end{figure*}

(ii) The chaotic dynamics occurs for the low-frequency part of the Fig.~\ref{fig:core:homega} and in the stretched region between resonances at $\omega_0$ and $2\omega_0$. An example of the temporal evolution is shown in the Fig.~\ref{fig:core:homega}c. The projection of the phase diagram on the $(\mu,\psi)$ plane (see Fig.~\ref{fig:core:homega}d) looks similarly to the pseudo--phase diagram in the Fig.~\ref{fig:sim:pseudofield}(a): the projection of trajectory is not closed and representation point makes a lot of windings around both $\pm \mu_0$. The shape of the chaotic Poincar\'{e} maps depends on the frequency. They show the shape of strange attractors, see Fig.~\ref{fig:core:poincareEvolution}. Their Cantor structure is ill--defined due to using low damping. Note, that they are similar to strange attractors for the Duffing oscillator\cite{Moon04} (nonlinear oscillator with quadratic and cubic nonlinearities in the double-well potential). However, the reduced core model has a more complicated
nonlinearity term; using the mechanical analogy one can speak about the motion of a particle with a variable mass (7) in a double--well potential.

(iii) The main part of the diagram of switching events~\ref{fig:core:homega} is 
occupied by the region of regular dynamics. The most frequently observed 
Poincar\'{e} maps for this case contain some number of stable focuses. The 
observed numbers are $\overline{1,12}$, 15, 16, 18, 21, 24, 30 and 96. The most 
frequently observed ones are 1 (the grey region in the 
Fig.~\ref{fig:core:homega}) and 3 (included into the blue region in the 
Fig.~\ref{fig:core:homega}). Some of the points with a higher number of focuses 
demonstrate a complicated regular dynamics in phase space, see 
Fig.~\ref{fig:core:homega}e. The analogue of quasi--rectangular regular polarity 
oscillations in OOMMF simulations is found in the $\omega_0/3$ region, see phase 
diagram in the Fig.~\ref{fig:core:homega}a and temporal evolution in the 
Fig.~\ref{fig:core:homega}b (compare with the pseudo--phase diagram shown in the 
Fig.~\ref{fig:sim:pseudofield}(b) and the temporal evolution in the 
Fig.~\ref{fig:sim:homega}d).

(iv) The analogue of the intermittent switching linked with perturbation by the applying of the external field are shown by two dark--grey color intensities in the Fig.~\ref{fig:core:homega}. The final dynamics is an oscillation around upward or downward polarity (points $(\mu_0,0)$ and $(-\mu_0,0)$ in the phase space projection, respectively). An example of the temporal evolution is shown in the Fig.~\ref{fig:core:homega}f. As in OOMMF simulations, such oscillations typically occur near the border of the switching region. As in case of the chaotic dynamics, the resulting polarity is highly dependent on field, frequency and integration conditions.

\section{Controlled switching}
\label{sec:control}

\begin{figure*}
\begin{center}
\begin{tabular}{cc}
\multirow{2}{*}[41pt]{%
\begin{subfigure}{0.48\textwidth}
\includegraphics[width=\columnwidth]{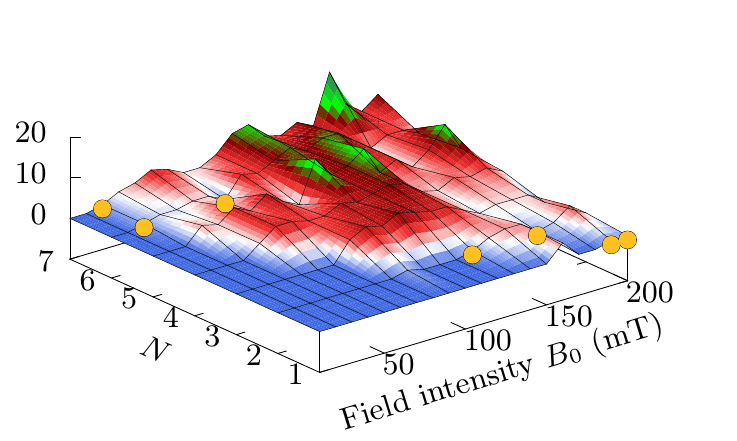}
\caption{Number of switchings.}
\label{fig:train}%
\end{subfigure}%
} & %
\begin{subfigure}{0.48\textwidth}
\includegraphics[width=\columnwidth]{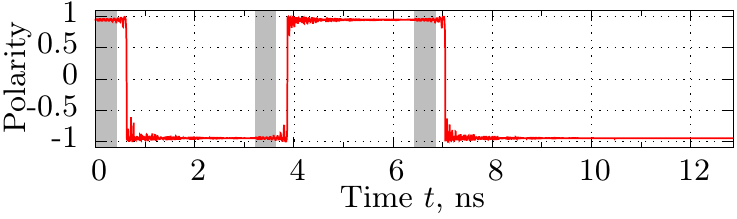}
\caption{OOMMF simulations}
\label{fig:unidirectional}%
\end{subfigure} \\
& %
\begin{subfigure}{0.48\textwidth}
\includegraphics[width=\columnwidth]{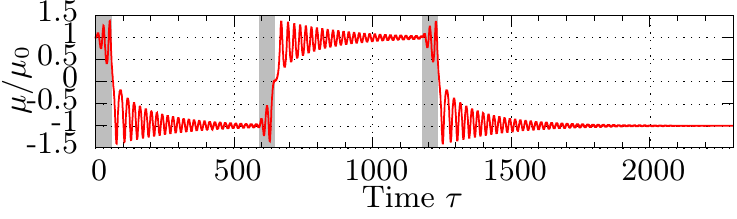}
\caption{Core model}
\label{fig:unidirectionalCore}%
\end{subfigure} \\
\end{tabular}
\end{center}
\caption{(Color online) \subref{fig:train} Number of switchings as function of number of oscillations in the wave train ($N$) and the field intensity ($B_0$). Symbols correspond to values $N$ and $B_0$, where unidirectional switching is observed. \subref{fig:unidirectional} Controlled unidirectional switching by sequence of short wave trains (contained $N=6$ periods on frequency 14\,GHz, duration 0.43\,ns) with period 3.2\,ns. Gray regions show time of applying external magnetic field. \subref{fig:unidirectionalCore} The same for core model. Separate pulse contains $N=3$ periods (duration $58.9$ in arb.~u.), interval between pulses $589$ arb.~units.}
\label{fig:sim:train}
\end{figure*}

As it is shown by analysis of the diagram of switchings events~(Fig.~\ref{fig:sim:homega}), the vortex polarity switching under the action of the perpendicular resonant field produces multiple switchings during a short time comparable with one period of the acting field. Such a situation is not unique among different polarity switching methods. So, for the axially--asymmetric scenario, sufficient strength of gaussian pulse in the sample's plane produces more than one sequential vortex--antivortex pair creation and annihilations.\cite{Hertel07} Using an in--plane rotating field with frequency $\omega_r$ codirectional with the vortex polarity $p$ ($\omega_r p > 0$) stabilizes the vortex in the center of the sample. But when $\omega_r p < 0$ the reversal occurs in a specific range of the field intensities and frequencies, above which multiple switching was observed. \cite{Kravchuk07c} A similar picture was reported in Ref.~\onlinecite{Liu07a} for the current--induced switching.

However for the further application in contrast to multiple switching a unidirectional vortex polarity reversal, is needed. Because the pumping~\eqref{eq:sim:Field} does not select any direction of the vortex polarity, the most natural way to avoid the multiple reversal consists in limiting the pulse duration. We test an influence of a short wave train in the form
\begin{equation} \label{eq:sim:trainField}
\vec B =
\begin{cases}
B_0\vec e_z \sin \left( 2\pi f t\right),& t \in \left[0,\dfrac{N}{f}\right], \\
0, & \text{otherwise},
\end{cases}
\end{equation}
where $N\in\mathbb{N}$ is the number of periods of the sinusoidal magnetic field in the wave train. We investigate the vortex dynamics under the action of the field \eqref{eq:sim:trainField} on the resonant frequency $f = f_0$. The vortex polarity is observed during a long time $10N/f$ in order to damp magnons.

The response of the magnetization to the field \eqref{eq:sim:trainField} shows the nonlinear dependency on $B_0$ and $N$, see~Fig.~\ref{fig:train}. When field intensity and period numbers are small, there are only small oscillations of the magnetization inside the vortex core. When $N$ becomes larger than some critical value, a typical system behaviour looks like a few switchings, which also occurs when the field is already turned off. A unidirectional switching from upward polarity to downward is observed for $B_0 = 30$\,mT and $N=6,7$ and some higher fields. We check the controllability of the discussed switching method by applying of series of wave trains. The series of wave trains is applied to the relaxed vortex. The time interval between trains is varied with steps of $0.5/f_0$ in different series. For $B_0=30$\,mT and $N=6$ the first switching occurs in 627\,ps and the vortex starts to relax (field is turned off at the time 429\,ps). The sequence containing 3 wave trains allows to get controllable
unidirectional vortex polarity reversal with minimal interval between wave trains of $3.2$\,ns, see Fig.~\ref{fig:unidirectional}. These time intervals correspond speed of changing state of such memory cell about 250\,MHz. The core model also gives the same qualitative result, see Fig.~\ref{fig:unidirectionalCore}.

\section{Discussion}
\label{sec:conclusion} %

The axially--symmetric switching of the vortex core under the action of periodic pumping was very recenltly predicted in Refs.~\onlinecite{Wang12,Yoo12}. \citet{Wang12} were concerned with switching events: for the typical nanodot size the switching at the resonant frequency occurs during 600\,ps. \citet{Yoo12} computed the diagram of switching events where they noticed the existence of resonances on double and triple harmonics and shown that the exchange energy becomes higher than the threshold value for the vortex core reversal. In this work we study the long-- and short--time vortex dynamics and propose an analytical model which describes the phenomena of full--scale simulations.

In order to explain the complicated vortex dynamics, we use here the reduced core model. \cite{Wysin94,Gaididei99,Gaididei00,Zagorodny03} It should be noted that this model does not pretend a quantitative agreement with simulations. In particular, it does not provide even the eigenfrequencies of the radially symmetric magnon modes, which is rather complicated task. \cite{Zaspel09, Galkin09} Nevertheless, the model we use is the simplest one which allows to describe a rich variety of different regimes of vortex polarity dynamics. This model provides a simple physical picture of the switching phenomenon in terms of the nonlinear resonance in a double--well potential. Such a potential arises mainly from the exchange interaction: the presence of two wells corresponds to the energy degeneracy with respect to the direction of the vortex polarity (up or down); the energy barrier between the wells becomes higher as the discreteness effects become less important. One has to stress that the switching process is 
forbidden in the continuum theory. In a real magnet the magnetization reversal is possible due to the discreteness of the lattice. That is why to describe the switching analytically we use the discrete core model: the switching can be considered as the motion of an effective mechanical particle with a variable mass in the double--well potential. Under the action of periodical pumping the particle starts to oscillate near the bottom of one of the wells. When the pumping increases, there appear nonlinear oscillations of the particle; under a further forcing the particle overcomes the barrier, which corresponds to the magnetization reversal process.

The chaotic dynamics of the magnetization is studied for domains walls\cite{Shutyi07} and current--induced phenomena in monodomain nanoparticles.\cite{Lee04a,Berkov05,Yang07a} Very recently the existence of incommensurate chaotic vortex dynamics in spin valves was reported.\cite{Petit-Watelot12} In our case the chaos enters in the vortex polarity switching process due to the periodical pumping of the system with two equivalent equilibrium states as it happens in a Duffing oscillator.\cite{Moon04}

The periodic pumping does not select the preferable vortex polarity direction which causes multiple switchings under the action of sufficiently high fields and frequencies. However, accurate fitting of the pulse duration and the time interval between sequential pulses allows to obtain a controlled unidirectional core reversal. Thus, the chaotic, regular and controlled vortex polarity dynamics could find applications in physical layer data encryption\cite{Kocarev95,Schoell08} and memory devices.\cite{Kim08,Yu11a}

In the current study we do not consider thermal effects on the vortex dynamics. The influence of the temperature was found to be not essential for the current--induced motion of an individual vortex in Py nanodisks \cite{Ishida06}. Nevertheless it should be noted while magnetic vortices are stable up to very high temperature,\cite{Muxworthy03} the heating can influence the gyroscopical vortex dynamics \cite{Kamionka11,Depondt12}. The heat can induce the vortex dynamics in the system. \cite{Machado12} The temperature--induced vortex dynamics also can influence the critical fields for vortex nucleation and annihilation.\cite{Mihajlovic10} We expect that the physical picture discussed in the paper with a variety of different dynamical regimes survives with the temperature. The thermal effects will cause the shift of boundaries between different regimes.

\begin{acknowledgments}

O.V.P. and D.D.S. thank the University of Bayreuth and Computing Center of the University of Bayreuth, where a part of this work was performed, for kind hospitality. O.V.P. acknowledges the support from the BAYHOST project. D.D.S. acknowledges the support from the Alexander von Humboldt Foundation. F.G.M. acknowledges the support by MICINN through FIS2011-24540, and by Junta de Andalucia under Project No. FQM207.  All simulations results presented in the work were obtained
using the computing cluster of Kiev University \cite{unicc} and Bayreuth University \cite{btrzx}.

\end{acknowledgments}

\appendix

\section{Reduced Core Model}
\label{app:core}

Taking into account the explicit form of the magnetic field, \eqref{eq:sim:Field}, the energy \eqref{eq:Energy} reads\cite{Zagorodny03}
\begin{equation} \label{eq:Energy-1}
\begin{split}
E &= -  \frac{A L_z}{2}\!\!\!\sum_{(\vec{n},\vec{\delta})}\! \Biggl[\!
\sqrt{(1-m_{\vec{n}}^2)(1-m_{\vec{n} +\vec{\delta}}^2)}\cos(\phi_{\vec{n}}\!
-\phi_{\vec{n}+\vec{\delta}})\\
& + \lambda m_{\vec{n}} m_{\vec{n}+\vec{\delta}}  \Bigr]- a_0^2 M_s L_z B_0
\sin\left(2\pi f t\right) \sum_{\vec{n}} m_{\vec{n}}.
\end{split}
\end{equation}
Now we incorporate here the reduce core Ansatz \eqref{eq:vortex}. Then the energy \eqref{eq:Energy-1} reads
\begin{equation} \label{eq:Energy-2}
\begin{split}
E &= -4a_0^2M_sL_z \mu B_0\sin(2\pi f t) - 4AL_z\lambda \mu^2 - \\
  & - \frac{16}{\sqrt{5}}AL_z\sqrt{1-\mu^2}\cos\psi.
\end{split}
\end{equation}
After the renormalization the Eq.~\ref{eq:Energy-2} takes the form~\eqref{eq:E-core}, where $ \mathscr{E} = E/\epsilon$, $\epsilon = 8AL_z\lambda$.

The magnetization dynamics follows the Landau--Lifshitz--Gilbert equations
\begin{equation} \label{eq:LL}
\frac{\mathrm d\vec{m}_{\vec{n}}}{\mathrm d\tau} = \vec{m}_{\vec{n}}\times
\frac{\partial \mathscr{E}}{\partial \vec{m}_{\vec{n}}} + \eta \vec{m}_{\vec{n}}
\times \frac{\mathrm d\vec{m}_{\vec{n}}}{\mathrm d\tau},
\end{equation}
with $\eta$ being a Gilbert damping coefficient and the rescaled time $\tau = \epsilon\gamma t/M_s$. Substituting Eqs.~\eqref{eq:magnMomTurnPh} into the Eq.~\eqref{eq:LL} we obtain the equations for the $(\mu,\psi)$:
\begin{equation}\label{eq:LLGnew}
\begin{aligned}
	\frac{\mathrm d\mu}{\mathrm d\tau} & = \frac{\partial \mathscr E}{\partial
\psi} - \eta (1-\mu^2) \frac{\partial \mathscr E}{\partial \mu},\\
	\frac{\mathrm d\psi}{\mathrm d\tau} & = - \frac{\partial \mathscr
E}{\partial \mu} - \frac{\eta}{1-\mu^2}	\frac{\partial \mathscr E}{\partial
\psi},
\end{aligned}
\end{equation}
Substituting Eq.~\eqref{eq:E-core} into \eqref{eq:LLGnew} we get the
Eq.~\eqref{eq:core:motion}.

\section{Weakly nonlinear analysis}
\label{app:wna}

Let us consider the weakly nonlinear case for Eqs.~\eqref{eq:core:motion}. Using the series expansion \eqref{eq:core:smallAmp}, the time derivative reads
\begin{equation}
\frac{\mathrm{d}}{\mathrm{d}t} = \sum_{n=0}^2\varepsilon^n D_n,\quad D_n =
\frac{\mathrm{d}}{\mathrm{d}T_n},
\end{equation}
and the equations of motion~\eqref{eq:core:motion} break into three pairs of equations for the different orders in $\varepsilon$:
\begin{align}
\label{eq:app:firm}
D_0 \mu_1 & = \Lambda^2 \psi_1,\\
\label{eq:app:firpsi}
D_0 \psi_1 & = \left(1-\frac{1}{\Lambda ^2}\right) \mu_1,\\
\label{eq:app:secm}
D_1 \mu_1 & + D_0 \mu_2 = - \sqrt{1-\Lambda ^2} \mu_1 \psi_1 + \Lambda ^2
\psi_2,\\
\label{eq:app:secpsi}
D_1 \psi_1 & + D_0 \psi_2 = \left(1-\frac{1}{\Lambda ^2}\right) \mu_2 - \frac{3
\sqrt{1-\Lambda ^2}}{2 \Lambda ^4} \mu_1^2 \\
 & + \frac{\sqrt{1-\Lambda ^2}}{2} \psi_1^2, \nonumber \\
\label{eq:app:thirm}
D_2 \mu_1 & + D_1 \mu_2 + D_0 \mu_3 = \Lambda^2 \psi_3 - \frac{1}{2 \Lambda ^2}
\mu_1^2\psi_1\\
 & - \sqrt{1-\Lambda ^2} (\mu_2\psi_1 + \mu_1 \psi_2) - \frac{\Lambda ^2}{6}
\psi_1^3 \nonumber\\
 & -\left(1-\Lambda ^2\right) \eta_2 \mu_1,\nonumber  \\
\label{eq:app:thirpsi}
D_2 \psi_1 & + D_1 \psi_2 + D_0 \psi_3 = \left(1-\frac{1}{\Lambda ^2}\right)
\mu_3\\
& - \frac{5-4 \Lambda ^2}{2 \Lambda ^6} \mu_1^3 - \frac{3 \sqrt{1-\Lambda
^2}}{\Lambda ^4} \mu_1 \mu_2 + \frac{1}{2 \Lambda ^2} \mu_1 \psi_1^2 \nonumber
\\
& +\sqrt{1-\Lambda ^2} \psi_1\psi_2 + h_3\sin \omega t - \eta_2 \psi_1.
\nonumber
\end{align}

The solution of Eqs.~\eqref{eq:app:firm} and \eqref{eq:app:firpsi} reads
\begin{equation*}
\mu_1(T_0, T_1, T_2) = A(T_1, T_2) e^{i \omega_0 T_0} + A^*(T_1, T_2) e^{-i
\omega_0 T_0}.
\end{equation*}
Following the Floquet theory \cite{Nayfeh85}, one needs to omit all secular terms. Thus, Eqs.~\eqref{eq:app:secm} and \eqref{eq:app:secpsi} show $A(T_1,T_2) \equiv A(T_2)$ and  Eqs.~\eqref{eq:app:thirm} and \eqref{eq:app:thirpsi} gives the equation for $A(T_2)$:
\begin{equation} \label{eq:app:lasteq}
\begin{split}
2 i \sqrt{1-\Lambda ^2} A' & + i \eta_2 \sqrt{1-\Lambda ^2} \left(2-\Lambda
^2\right) A \\
& - 2\frac{2+\Lambda^2}{\Lambda^2} A^2 A^* = \Lambda^2
\frac{h_3}{2i}e^{i\omega_2 T_2}.
\end{split}
\end{equation}
By solving the Eq.~\eqref{eq:app:lasteq} one obtains the Eq.~\eqref{eq:core:resonanceCurve}.


\begin{thebibliography}{51}%
\makeatletter
\providecommand \@ifxundefined [1]{%
 \@ifx{#1\undefined}
}%
\providecommand \@ifnum [1]{%
 \ifnum #1\expandafter \@firstoftwo
 \else \expandafter \@secondoftwo
 \fi
}%
\providecommand \@ifx [1]{%
 \ifx #1\expandafter \@firstoftwo
 \else \expandafter \@secondoftwo
 \fi
}%
\providecommand \natexlab [1]{#1}%
\providecommand \enquote  [1]{``#1''}%
\providecommand \bibnamefont  [1]{#1}%
\providecommand \bibfnamefont [1]{#1}%
\providecommand \citenamefont [1]{#1}%
\providecommand \href@noop [0]{\@secondoftwo}%
\providecommand \href [0]{\begingroup \@sanitize@url \@href}%
\providecommand \@href[1]{\@@startlink{#1}\@@href}%
\providecommand \@@href[1]{\endgroup#1\@@endlink}%
\providecommand \@sanitize@url [0]{\catcode `\\12\catcode `\$12\catcode
  `\&12\catcode `\#12\catcode `\^12\catcode `\_12\catcode `\%12\relax}%
\providecommand \@@startlink[1]{}%
\providecommand \@@endlink[0]{}%
\providecommand \url  [0]{\begingroup\@sanitize@url \@url }%
\providecommand \@url [1]{\endgroup\@href {#1}{\urlprefix }}%
\providecommand \urlprefix  [0]{URL }%
\providecommand \Eprint [0]{\href }%
\providecommand \doibase [0]{http://dx.doi.org/}%
\providecommand \selectlanguage [0]{\@gobble}%
\providecommand \bibinfo  [0]{\@secondoftwo}%
\providecommand \bibfield  [0]{\@secondoftwo}%
\providecommand \translation [1]{[#1]}%
\providecommand \BibitemOpen [0]{}%
\providecommand \bibitemStop [0]{}%
\providecommand \bibitemNoStop [0]{.\EOS\space}%
\providecommand \EOS [0]{\spacefactor3000\relax}%
\providecommand \BibitemShut  [1]{\csname bibitem#1\endcsname}%
\let\auto@bib@innerbib\@empty
\bibitem [{\citenamefont {Braun}(2012)}]{Braun12}%
  \BibitemOpen
  \bibfield  {author} {\bibinfo {author} {\bibfnamefont {H.-B.}\ \bibnamefont
  {Braun}},\ }\href {\doibase 10.1080/00018732.2012.663070} {\bibfield
  {journal} {\bibinfo  {journal} {Advances in Physics}\ }\textbf {\bibinfo
  {volume} {61}},\ \bibinfo {pages} {1} (\bibinfo {year} {2012})}\BibitemShut
  {NoStop}%
\bibitem [{\citenamefont {Hubert}\ and\ \citenamefont {Sch{\"
  a}fer}(1998)}]{Hubert98}%
  \BibitemOpen
  \bibfield  {author} {\bibinfo {author} {\bibfnamefont {A.}~\bibnamefont
  {Hubert}}\ and\ \bibinfo {author} {\bibfnamefont {R.}~\bibnamefont {Sch{\"
  a}fer}},\ }\href@noop {} {\emph {\bibinfo {title} {Magnetic domains: the
  analysis of magnetic microstructures}}}\ (\bibinfo  {publisher}
  {Springer--Verlag},\ \bibinfo {address} {Berlin},\ \bibinfo {year}
  {1998})\BibitemShut {NoStop}%
\bibitem [{\citenamefont {Wachowiak}\ \emph {et~al.}(2002)\citenamefont
  {Wachowiak}, \citenamefont {Wiebe}, \citenamefont {Bode}, \citenamefont
  {Pietzsch}, \citenamefont {Morgenstern},\ and\ \citenamefont
  {Wiesendanger}}]{Wachowiak02}%
  \BibitemOpen
  \bibfield  {author} {\bibinfo {author} {\bibfnamefont {A.}~\bibnamefont
  {Wachowiak}}, \bibinfo {author} {\bibfnamefont {J.}~\bibnamefont {Wiebe}},
  \bibinfo {author} {\bibfnamefont {M.}~\bibnamefont {Bode}}, \bibinfo {author}
  {\bibfnamefont {O.}~\bibnamefont {Pietzsch}}, \bibinfo {author}
  {\bibfnamefont {M.}~\bibnamefont {Morgenstern}}, \ and\ \bibinfo {author}
  {\bibfnamefont {R.}~\bibnamefont {Wiesendanger}},\ }\href {\doibase
  10.1126/science.1075302} {\bibfield  {journal} {\bibinfo  {journal}
  {science}\ }\textbf {\bibinfo {volume} {298}},\ \bibinfo {pages} {577}
  (\bibinfo {year} {2002})}\BibitemShut {NoStop}%
\bibitem [{\citenamefont {Kim}\ \emph {et~al.}(2008)\citenamefont {Kim},
  \citenamefont {Lee}, \citenamefont {Yu},\ and\ \citenamefont {Choi}}]{Kim08}%
  \BibitemOpen
  \bibfield  {author} {\bibinfo {author} {\bibfnamefont {S.-K.}\ \bibnamefont
  {Kim}}, \bibinfo {author} {\bibfnamefont {K.-S.}\ \bibnamefont {Lee}},
  \bibinfo {author} {\bibfnamefont {Y.-S.}\ \bibnamefont {Yu}}, \ and\ \bibinfo
  {author} {\bibfnamefont {Y.-S.}\ \bibnamefont {Choi}},\ }\href {\doibase
  10.1063/1.2807274} {\bibfield  {journal} {\bibinfo  {journal} {Appl. Phys.
  Lett.}\ }\textbf {\bibinfo {volume} {92}},\ \bibinfo {eid} {022509} (\bibinfo
  {year} {2008})}\BibitemShut {NoStop}%
\bibitem [{\citenamefont {Yu}\ \emph {et~al.}(2011)\citenamefont {Yu},
  \citenamefont {Jung}, \citenamefont {Lee}, \citenamefont {Fischer},\ and\
  \citenamefont {Kim}}]{Yu11a}%
  \BibitemOpen
  \bibfield  {author} {\bibinfo {author} {\bibfnamefont {Y.-S.}\ \bibnamefont
  {Yu}}, \bibinfo {author} {\bibfnamefont {H.}~\bibnamefont {Jung}}, \bibinfo
  {author} {\bibfnamefont {K.-S.}\ \bibnamefont {Lee}}, \bibinfo {author}
  {\bibfnamefont {P.}~\bibnamefont {Fischer}}, \ and\ \bibinfo {author}
  {\bibfnamefont {S.-K.}\ \bibnamefont {Kim}},\ }\href {\doibase
  10.1063/1.3551524} {\bibfield  {journal} {\bibinfo  {journal} {Appl. Phys.
  Lett.}\ }\textbf {\bibinfo {volume} {98}},\ \bibinfo {eid} {052507} (\bibinfo
  {year} {2011})}\BibitemShut {NoStop}%
\bibitem [{\citenamefont {Gaididei}\ \emph {et~al.}(1999)\citenamefont
  {Gaididei}, \citenamefont {Kamppeter}, \citenamefont {Mertens},\ and\
  \citenamefont {Bishop}}]{Gaididei99}%
  \BibitemOpen
  \bibfield  {author} {\bibinfo {author} {\bibfnamefont {Y.}~\bibnamefont
  {Gaididei}}, \bibinfo {author} {\bibfnamefont {T.}~\bibnamefont {Kamppeter}},
  \bibinfo {author} {\bibfnamefont {F.~G.}\ \bibnamefont {Mertens}}, \ and\
  \bibinfo {author} {\bibfnamefont {A.}~\bibnamefont {Bishop}},\ }\href
  {http://link.aps.org/abstract/PRB/v59/p7010} {\bibfield  {journal} {\bibinfo
  {journal} {Phys. Rev. B}\ }\textbf {\bibinfo {volume} {59}},\ \bibinfo
  {pages} {7010} (\bibinfo {year} {1999})}\BibitemShut {NoStop}%
\bibitem [{\citenamefont {Gaididei}\ \emph {et~al.}(2000)\citenamefont
  {Gaididei}, \citenamefont {Kamppeter}, \citenamefont {Mertens},\ and\
  \citenamefont {Bishop}}]{Gaididei00}%
  \BibitemOpen
  \bibfield  {author} {\bibinfo {author} {\bibfnamefont {Y.}~\bibnamefont
  {Gaididei}}, \bibinfo {author} {\bibfnamefont {T.}~\bibnamefont {Kamppeter}},
  \bibinfo {author} {\bibfnamefont {F.~G.}\ \bibnamefont {Mertens}}, \ and\
  \bibinfo {author} {\bibfnamefont {A.~R.}\ \bibnamefont {Bishop}},\ }\href
  {\doibase 10.1103/PhysRevB.61.9449} {\bibfield  {journal} {\bibinfo
  {journal} {Phys. Rev. B}\ }\textbf {\bibinfo {volume} {61}},\ \bibinfo
  {pages} {9449} (\bibinfo {year} {2000})}\BibitemShut {NoStop}%
\bibitem [{\citenamefont {Van~Waeyenberge}\ \emph {et~al.}(2006)\citenamefont
  {Van~Waeyenberge}, \citenamefont {Puzic}, \citenamefont {Stoll},
  \citenamefont {Chou}, \citenamefont {Tyliszczak}, \citenamefont {Hertel},
  \citenamefont {F\"ahnle}, \citenamefont {Bruckl}, \citenamefont {Rott},
  \citenamefont {Reiss}, \citenamefont {Neudecker}, \citenamefont {Weiss},
  \citenamefont {Back},\ and\ \citenamefont {Sch\"utz}}]{Waeyenberge06}%
  \BibitemOpen
  \bibfield  {author} {\bibinfo {author} {\bibfnamefont {B.}~\bibnamefont
  {Van~Waeyenberge}}, \bibinfo {author} {\bibfnamefont {A.}~\bibnamefont
  {Puzic}}, \bibinfo {author} {\bibfnamefont {H.}~\bibnamefont {Stoll}},
  \bibinfo {author} {\bibfnamefont {K.~W.}\ \bibnamefont {Chou}}, \bibinfo
  {author} {\bibfnamefont {T.}~\bibnamefont {Tyliszczak}}, \bibinfo {author}
  {\bibfnamefont {R.}~\bibnamefont {Hertel}}, \bibinfo {author} {\bibfnamefont
  {M.}~\bibnamefont {F\"ahnle}}, \bibinfo {author} {\bibfnamefont
  {H.}~\bibnamefont {Bruckl}}, \bibinfo {author} {\bibfnamefont
  {K.}~\bibnamefont {Rott}}, \bibinfo {author} {\bibfnamefont {G.}~\bibnamefont
  {Reiss}}, \bibinfo {author} {\bibfnamefont {I.}~\bibnamefont {Neudecker}},
  \bibinfo {author} {\bibfnamefont {D.}~\bibnamefont {Weiss}}, \bibinfo
  {author} {\bibfnamefont {C.~H.}\ \bibnamefont {Back}}, \ and\ \bibinfo
  {author} {\bibfnamefont {G.}~\bibnamefont {Sch\"utz}},\ }\href
  {http://dx.doi.org/10.1038/nature05240} {\bibfield  {journal} {\bibinfo
  {journal} {Nature}\ }\textbf {\bibinfo {volume} {444}},\ \bibinfo {pages}
  {461} (\bibinfo {year} {2006})}\BibitemShut {NoStop}%
\bibitem [{\citenamefont {Okuno}\ \emph {et~al.}(2002)\citenamefont {Okuno},
  \citenamefont {Shigeto}, \citenamefont {Ono}, \citenamefont {Mibu},\ and\
  \citenamefont {Shinjo}}]{Okuno02}%
  \BibitemOpen
  \bibfield  {author} {\bibinfo {author} {\bibfnamefont {T.}~\bibnamefont
  {Okuno}}, \bibinfo {author} {\bibfnamefont {K.}~\bibnamefont {Shigeto}},
  \bibinfo {author} {\bibfnamefont {T.}~\bibnamefont {Ono}}, \bibinfo {author}
  {\bibfnamefont {K.}~\bibnamefont {Mibu}}, \ and\ \bibinfo {author}
  {\bibfnamefont {T.}~\bibnamefont {Shinjo}},\ }\href
  {http://www.sciencedirect.com/science/article/B6TJJ-447DCMV-3/2/84e9fdfbf9e0cab3d3182fc4db4b4032}
  {\bibfield  {journal} {\bibinfo  {journal} {J.~Magn. Magn. Mater.}\ }\textbf
  {\bibinfo {volume} {240}},\ \bibinfo {pages} {1} (\bibinfo {year}
  {2002})}\BibitemShut {NoStop}%
\bibitem [{\citenamefont {Thiaville}\ \emph {et~al.}(2003)\citenamefont
  {Thiaville}, \citenamefont {Garcia}, \citenamefont {Dittrich}, \citenamefont
  {Miltat},\ and\ \citenamefont {Schrefl}}]{Thiaville03}%
  \BibitemOpen
  \bibfield  {author} {\bibinfo {author} {\bibfnamefont {A.}~\bibnamefont
  {Thiaville}}, \bibinfo {author} {\bibfnamefont {J.~M.}\ \bibnamefont
  {Garcia}}, \bibinfo {author} {\bibfnamefont {R.}~\bibnamefont {Dittrich}},
  \bibinfo {author} {\bibfnamefont {J.}~\bibnamefont {Miltat}}, \ and\ \bibinfo
  {author} {\bibfnamefont {T.}~\bibnamefont {Schrefl}},\ }\href
  {http://link.aps.org/abstract/PRB/v67/e094410} {\bibfield  {journal}
  {\bibinfo  {journal} {Phys. Rev. B}\ }\textbf {\bibinfo {volume} {67}},\
  \bibinfo {eid} {094410} (\bibinfo {year} {2003})}\BibitemShut {NoStop}%
\bibitem [{\citenamefont {Kravchuk}\ and\ \citenamefont
  {Sheka}(2007)}]{Kravchuk07a}%
  \BibitemOpen
  \bibfield  {author} {\bibinfo {author} {\bibfnamefont {V.}~\bibnamefont
  {Kravchuk}}\ and\ \bibinfo {author} {\bibfnamefont {D.}~\bibnamefont
  {Sheka}},\ }\href {http://dx.doi.org/10.1134/S1063783407100186} {\bibfield
  {journal} {\bibinfo  {journal} {Physics of the Solid State}\ }\textbf
  {\bibinfo {volume} {49}},\ \bibinfo {pages} {1923} (\bibinfo {year}
  {2007})}\BibitemShut {NoStop}%
\bibitem [{\citenamefont {Vila}\ \emph {et~al.}(2009)\citenamefont {Vila},
  \citenamefont {Darques}, \citenamefont {Encinas}, \citenamefont {Ebels},
  \citenamefont {George}, \citenamefont {Faini}, \citenamefont {Thiaville},\
  and\ \citenamefont {Piraux}}]{Vila09}%
  \BibitemOpen
  \bibfield  {author} {\bibinfo {author} {\bibfnamefont {L.}~\bibnamefont
  {Vila}}, \bibinfo {author} {\bibfnamefont {M.}~\bibnamefont {Darques}},
  \bibinfo {author} {\bibfnamefont {A.}~\bibnamefont {Encinas}}, \bibinfo
  {author} {\bibfnamefont {U.}~\bibnamefont {Ebels}}, \bibinfo {author}
  {\bibfnamefont {J.-M.}\ \bibnamefont {George}}, \bibinfo {author}
  {\bibfnamefont {G.}~\bibnamefont {Faini}}, \bibinfo {author} {\bibfnamefont
  {A.}~\bibnamefont {Thiaville}}, \ and\ \bibinfo {author} {\bibfnamefont
  {L.}~\bibnamefont {Piraux}},\ }\href {\doibase 10.1103/PhysRevB.79.172410}
  {\bibfield  {journal} {\bibinfo  {journal} {Phys. Rev. B}\ }\textbf {\bibinfo
  {volume} {79}},\ \bibinfo {eid} {172410} (\bibinfo {year}
  {2009})}\BibitemShut {NoStop}%
\bibitem [{\citenamefont {Kravchuk}\ \emph {et~al.}(2009)\citenamefont
  {Kravchuk}, \citenamefont {Gaididei},\ and\ \citenamefont
  {Sheka}}]{Kravchuk09}%
  \BibitemOpen
  \bibfield  {author} {\bibinfo {author} {\bibfnamefont {V.~P.}\ \bibnamefont
  {Kravchuk}}, \bibinfo {author} {\bibfnamefont {Y.}~\bibnamefont {Gaididei}},
  \ and\ \bibinfo {author} {\bibfnamefont {D.~D.}\ \bibnamefont {Sheka}},\
  }\href {\doibase 10.1103/PhysRevB.80.100405} {\bibfield  {journal} {\bibinfo
  {journal} {Phys. Rev. B}\ }\textbf {\bibinfo {volume} {80}},\ \bibinfo {eid}
  {100405} (\bibinfo {year} {2009})}\BibitemShut {NoStop}%
\bibitem [{\citenamefont {Gaididei}\ \emph {et~al.}(2010)\citenamefont
  {Gaididei}, \citenamefont {Kravchuk}, \citenamefont {Sheka},\ and\
  \citenamefont {Mertens}}]{Gaididei10b}%
  \BibitemOpen
  \bibfield  {author} {\bibinfo {author} {\bibfnamefont {Y.}~\bibnamefont
  {Gaididei}}, \bibinfo {author} {\bibfnamefont {V.~P.}\ \bibnamefont
  {Kravchuk}}, \bibinfo {author} {\bibfnamefont {D.~D.}\ \bibnamefont {Sheka}},
  \ and\ \bibinfo {author} {\bibfnamefont {F.~G.}\ \bibnamefont {Mertens}},\
  }\href {\doibase 10.1103/PhysRevB.81.094431} {\bibfield  {journal} {\bibinfo
  {journal} {Phys. Rev. B}\ }\textbf {\bibinfo {volume} {81}},\ \bibinfo
  {pages} {094431} (\bibinfo {year} {2010})}\BibitemShut {NoStop}%
\bibitem [{\citenamefont {Gaididei}\ \emph {et~al.}(2008)\citenamefont
  {Gaididei}, \citenamefont {Kravchuk}, \citenamefont {Sheka},\ and\
  \citenamefont {Mertens}}]{Gaididei08b}%
  \BibitemOpen
  \bibfield  {author} {\bibinfo {author} {\bibfnamefont {Y.~B.}\ \bibnamefont
  {Gaididei}}, \bibinfo {author} {\bibfnamefont {V.~P.}\ \bibnamefont
  {Kravchuk}}, \bibinfo {author} {\bibfnamefont {D.~D.}\ \bibnamefont {Sheka}},
  \ and\ \bibinfo {author} {\bibfnamefont {F.~G.}\ \bibnamefont {Mertens}},\
  }\href {\doibase 10.1063/1.2957013} {\bibfield  {journal} {\bibinfo
  {journal} {Low Temperature Physics}\ }\textbf {\bibinfo {volume} {34}},\
  \bibinfo {pages} {528} (\bibinfo {year} {2008})}\BibitemShut {NoStop}%
\bibitem [{\citenamefont {Wang}\ and\ \citenamefont {Dong}(2012)}]{Wang12}%
  \BibitemOpen
  \bibfield  {author} {\bibinfo {author} {\bibfnamefont {R.}~\bibnamefont
  {Wang}}\ and\ \bibinfo {author} {\bibfnamefont {X.}~\bibnamefont {Dong}},\
  }\href {\doibase 10.1063/1.3687909} {\bibfield  {journal} {\bibinfo
  {journal} {Appl. Phys. Lett.}\ }\textbf {\bibinfo {volume} {100}},\ \bibinfo
  {eid} {082402} (\bibinfo {year} {2012})}\BibitemShut {NoStop}%
\bibitem [{\citenamefont {Yoo}\ \emph {et~al.}(2012)\citenamefont {Yoo},
  \citenamefont {Lee},\ and\ \citenamefont {Kim}}]{Yoo12}%
  \BibitemOpen
  \bibfield  {author} {\bibinfo {author} {\bibfnamefont {M.-W.}\ \bibnamefont
  {Yoo}}, \bibinfo {author} {\bibfnamefont {J.}~\bibnamefont {Lee}}, \ and\
  \bibinfo {author} {\bibfnamefont {S.-K.}\ \bibnamefont {Kim}},\ }\href
  {\doibase 10.1063/1.4705690} {\bibfield  {journal} {\bibinfo  {journal}
  {Appl. Phys. Lett.}\ }\textbf {\bibinfo {volume} {100}},\ \bibinfo {eid}
  {172413} (\bibinfo {year} {2012})}\BibitemShut {NoStop}%
\bibitem [{\citenamefont {Pylypovskyi}\ \emph {et~al.}(2013)\citenamefont
  {Pylypovskyi}, \citenamefont {Sheka}, \citenamefont {Kravchuk}, \citenamefont
  {Gaididei},\ and\ \citenamefont {Mertens}}]{Pylypovskyi13b}%
  \BibitemOpen
  \bibfield  {author} {\bibinfo {author} {\bibfnamefont {O.~V.}\ \bibnamefont
  {Pylypovskyi}}, \bibinfo {author} {\bibfnamefont {D.~D.}\ \bibnamefont
  {Sheka}}, \bibinfo {author} {\bibfnamefont {V.~P.}\ \bibnamefont {Kravchuk}},
  \bibinfo {author} {\bibfnamefont {Y.}~\bibnamefont {Gaididei}}, \ and\
  \bibinfo {author} {\bibfnamefont {F.~G.}\ \bibnamefont {Mertens}},\ }\href
  {http://www.ujp.bitp.kiev.ua/files/journals/58/6/580611p.pdf} {\bibfield
  {journal} {\bibinfo  {journal} {Ukr. J. Phys.}\ }\textbf {\bibinfo {volume}
  {58}},\ \bibinfo {pages} {596} (\bibinfo {year} {2013})}\BibitemShut
  {NoStop}%
\bibitem [{\citenamefont {Zagorodny}\ \emph {et~al.}(2003)\citenamefont
  {Zagorodny}, \citenamefont {Gaididei}, \citenamefont {Mertens},\ and\
  \citenamefont {Bishop}}]{Zagorodny03}%
  \BibitemOpen
  \bibfield  {author} {\bibinfo {author} {\bibfnamefont {J.~P.}\ \bibnamefont
  {Zagorodny}}, \bibinfo {author} {\bibfnamefont {Y.}~\bibnamefont {Gaididei}},
  \bibinfo {author} {\bibfnamefont {F.~G.}\ \bibnamefont {Mertens}}, \ and\
  \bibinfo {author} {\bibfnamefont {A.~R.}\ \bibnamefont {Bishop}},\ }\href
  {\doibase 10.1140/epjb/e2003-00057-y} {\bibfield  {journal} {\bibinfo
  {journal} {Eur.~Phys.~J.}\ }\textbf {\bibinfo {volume} {B 31}},\ \bibinfo
  {pages} {471} (\bibinfo {year} {2003})}\BibitemShut {NoStop}%
\bibitem [{\citenamefont {Kim}(2010)}]{Kim10}%
  \BibitemOpen
  \bibfield  {author} {\bibinfo {author} {\bibfnamefont {S.-K.}\ \bibnamefont
  {Kim}},\ }\href {http://stacks.iop.org/0022-3727/43/i=26/a=264004} {\bibfield
   {journal} {\bibinfo  {journal} {Journal of Physics D: Applied Physics}\
  }\textbf {\bibinfo {volume} {43}},\ \bibinfo {pages} {264004} (\bibinfo
  {year} {2010})}\BibitemShut {NoStop}%
\bibitem [{OOM()}]{OOMMF}%
  \BibitemOpen
  \href {http://math.nist.gov/oommf/} {\enquote {\bibinfo {title} {The {O}bject
  {O}riented {M}icro{M}agnetic {F}ramework},}\ }\bibinfo {note} {Developed by
  M. J. Donahue and D. Porter mainly, from NIST. We used the 3D version of the
  1.2$\alpha$5 release}\BibitemShut {NoStop}%
\bibitem [{Note1()}]{Note1}%
  \BibitemOpen
  \bibinfo {note} {We also check the polarity dynamics for the samples with the
  same diameter and height 33\protect \tmspace +\thinmuskip {.1667em}nm, with
  diameter 360\protect \tmspace +\thinmuskip {.1667em}nm and heights 21\protect
  \tmspace +\thinmuskip {.1667em}nm and 33\protect \tmspace +\thinmuskip
  {.1667em}nm under the action of five different pairs of field intensity and
  frequency: (60\protect \tmspace +\thinmuskip {.1667em}mT, 10\protect \tmspace
  +\thinmuskip {.1667em}GHz), (60\protect \tmspace +\thinmuskip {.1667em}mT,
  13\protect \tmspace +\thinmuskip {.1667em}GHz), (80\protect \tmspace
  +\thinmuskip {.1667em}mT, 13\protect \tmspace +\thinmuskip {.1667em}GHz),
  (100\protect \tmspace +\thinmuskip {.1667em}mT, 13\protect \tmspace
  +\thinmuskip {.1667em}GHz), (100\protect \tmspace +\thinmuskip {.1667em}mT,
  18\protect \tmspace +\thinmuskip {.1667em}GHz). The qualitative behavior of
  the system described in this section remains the same. Nevertheless the
  geometrical parameters influence the frequencies of magnon modes and the
  threshold value of the switching field. Therefore, we expect that the change
  of the nanodot size shifts the characteristic frequencies of the minimal
  threshold fields and the threshold field amplitudes, while the qualitative
  behavior of the system remains the same.}\BibitemShut {Stop}%
\bibitem [{\citenamefont {Hertel}\ \emph {et~al.}(2007)\citenamefont {Hertel},
  \citenamefont {Gliga}, \citenamefont {F\"ahnle},\ and\ \citenamefont
  {Schneider}}]{Hertel07}%
  \BibitemOpen
  \bibfield  {author} {\bibinfo {author} {\bibfnamefont {R.}~\bibnamefont
  {Hertel}}, \bibinfo {author} {\bibfnamefont {S.}~\bibnamefont {Gliga}},
  \bibinfo {author} {\bibfnamefont {M.}~\bibnamefont {F\"ahnle}}, \ and\
  \bibinfo {author} {\bibfnamefont {C.~M.}\ \bibnamefont {Schneider}},\ }\href
  {http://link.aps.org/abstract/PRL/v98/e117201} {\bibfield  {journal}
  {\bibinfo  {journal} {Phys. Rev. Lett.}\ }\textbf {\bibinfo {volume} {98}},\
  \bibinfo {eid} {117201} (\bibinfo {year} {2007})}\BibitemShut {NoStop}%
\bibitem [{Note2()}]{Note2}%
  \BibitemOpen
  \bibinfo {note} {The field and frequency increments are varied depending on
  the position on the diagram of switching events: 10\protect \tmspace
  +\thinmuskip {.1667em}mT and 1\protect \tmspace +\thinmuskip {.1667em}GHz in
  the range $9\div 19$\protect \tmspace +\thinmuskip {.1667em}GHz and $10\div
  180$\protect \tmspace +\thinmuskip {.1667em}mT, 10\protect \tmspace
  +\thinmuskip {.1667em}mT and 0.5\protect \tmspace +\thinmuskip {.1667em}GHz
  in the range $3\div 8.5$\protect \tmspace +\thinmuskip {.1667em}GHz and
  $100\div 500$\protect \tmspace +\thinmuskip {.1667em}mT along the border of
  the diagram, 50\protect \tmspace +\thinmuskip {.1667em}mT and 1\protect
  \tmspace +\thinmuskip {.1667em}GHz for other ranges. For the frequencies $f >
  8.5$\protect \tmspace +\thinmuskip {.1667em}GHz the full time of simulations
  was 10\protect \tmspace +\thinmuskip {.1667em}ns [extended till 30\protect
  \tmspace +\thinmuskip {.1667em}ns for some particular points $(f,B_0)$] and
  for frequencies $f \le 8.5$\protect \tmspace +\thinmuskip {.1667em}GHz the
  full time of simulations was 5\protect \tmspace +\thinmuskip
  {.1667em}ns.}\BibitemShut {Stop}%
\bibitem [{Note3()}]{Note3}%
  \BibitemOpen
  \bibinfo {note} {Note that the vortex polarity instability which correspond
  to the nonlinear resonance was predicted in our previous paper, see
  Ref.~\protect \rev@citealpnum {Pylypovskyi13b}.}\BibitemShut {Stop}%
\bibitem [{\citenamefont {Moon}(2004)}]{Moon04}%
  \BibitemOpen
  \bibfield  {author} {\bibinfo {author} {\bibfnamefont {F.~C.}\ \bibnamefont
  {Moon}},\ }\href@noop {} {\emph {\bibinfo {title} {Chaotic Vibrations}}}\
  (\bibinfo  {publisher} {Wiley-VCH},\ \bibinfo {year} {2004})\ p.\ \bibinfo
  {pages} {309}\BibitemShut {NoStop}%
\bibitem [{Note4()}]{Note4}%
  \BibitemOpen
  \bibinfo {note} {As opposed frequency--independent white noise, the pink
  noise is characterized by the power law decay with $1/f^\beta $ spectrum,
  where $\beta \in (0,2)$, see Ref.~\protect \rev@citealpnum
  [p.~102]{Chen04}.}\BibitemShut {Stop}%
\bibitem [{\citenamefont {Wysin}(1994)}]{Wysin94}%
  \BibitemOpen
  \bibfield  {author} {\bibinfo {author} {\bibfnamefont {G.~M.}\ \bibnamefont
  {Wysin}},\ }\href {\doibase 10.1103/PhysRevB.49.8780} {\bibfield  {journal}
  {\bibinfo  {journal} {Phys. Rev. B}\ }\textbf {\bibinfo {volume} {49}},\
  \bibinfo {pages} {8780} (\bibinfo {year} {1994})}\BibitemShut {NoStop}%
\bibitem [{\citenamefont {Kevorkian}\ and\ \citenamefont
  {Cole}(1981)}]{Kevorkian81}%
  \BibitemOpen
  \bibfield  {author} {\bibinfo {author} {\bibfnamefont {J.}~\bibnamefont
  {Kevorkian}}\ and\ \bibinfo {author} {\bibfnamefont {J.}~\bibnamefont
  {Cole}},\ }\href {http://books.google.com.ua/books?id=pT3vAAAAMAAJ} {\emph
  {\bibinfo {title} {Perturbation methods in applied mathematics}}},\ Applied
  mathematical sciences\ (\bibinfo  {publisher} {Springer-Verlag},\ \bibinfo
  {year} {1981})\BibitemShut {NoStop}%
\bibitem [{\citenamefont {Nayfeh}(1985)}]{Nayfeh85}%
  \BibitemOpen
  \bibfield  {author} {\bibinfo {author} {\bibfnamefont {A.}~\bibnamefont
  {Nayfeh}},\ }\href {http://books.google.com.ua/books?id=3zbvAAAAMAAJ} {\emph
  {\bibinfo {title} {Problems in perturbation}}},\ A Wiley Interscience
  publication\ (\bibinfo  {publisher} {Wiley},\ \bibinfo {year}
  {1985})\BibitemShut {NoStop}%
\bibitem [{\citenamefont {Nayfeh}(2008)}]{Nayfeh08}%
  \BibitemOpen
  \bibfield  {author} {\bibinfo {author} {\bibfnamefont {A.}~\bibnamefont
  {Nayfeh}},\ }\href {http://books.google.com.ua/books?id=eh6RmWZ51NIC} {\emph
  {\bibinfo {title} {Perturbation Methods}}},\ Physics textbook\ (\bibinfo
  {publisher} {John Wiley \& Sons},\ \bibinfo {year} {2008})\BibitemShut
  {NoStop}%
\bibitem [{\citenamefont {Kravchuk}\ \emph {et~al.}(2007)\citenamefont
  {Kravchuk}, \citenamefont {Sheka}, \citenamefont {Gaididei},\ and\
  \citenamefont {Mertens}}]{Kravchuk07c}%
  \BibitemOpen
  \bibfield  {author} {\bibinfo {author} {\bibfnamefont {V.~P.}\ \bibnamefont
  {Kravchuk}}, \bibinfo {author} {\bibfnamefont {D.~D.}\ \bibnamefont {Sheka}},
  \bibinfo {author} {\bibfnamefont {Y.}~\bibnamefont {Gaididei}}, \ and\
  \bibinfo {author} {\bibfnamefont {F.~G.}\ \bibnamefont {Mertens}},\ }\href
  {\doibase 10.1063/1.2770819} {\bibfield  {journal} {\bibinfo  {journal}
  {J.~Appl. Phys.}\ }\textbf {\bibinfo {volume} {102}},\ \bibinfo {eid}
  {043908} (\bibinfo {year} {2007})}\BibitemShut {NoStop}%
\bibitem [{\citenamefont {Liu}\ \emph {et~al.}(2007)\citenamefont {Liu},
  \citenamefont {Gliga}, \citenamefont {Hertel},\ and\ \citenamefont
  {Schneider}}]{Liu07a}%
  \BibitemOpen
  \bibfield  {author} {\bibinfo {author} {\bibfnamefont {Y.}~\bibnamefont
  {Liu}}, \bibinfo {author} {\bibfnamefont {S.}~\bibnamefont {Gliga}}, \bibinfo
  {author} {\bibfnamefont {R.}~\bibnamefont {Hertel}}, \ and\ \bibinfo {author}
  {\bibfnamefont {C.~M.}\ \bibnamefont {Schneider}},\ }\href {\doibase
  10.1063/1.2780107} {\bibfield  {journal} {\bibinfo  {journal} {Appl. Phys.
  Lett.}\ }\textbf {\bibinfo {volume} {91}},\ \bibinfo {eid} {112501} (\bibinfo
  {year} {2007})}\BibitemShut {NoStop}%
\bibitem [{\citenamefont {Zaspel}\ \emph {et~al.}(2009)\citenamefont {Zaspel},
  \citenamefont {Wright}, \citenamefont {Galkin},\ and\ \citenamefont
  {Ivanov}}]{Zaspel09}%
  \BibitemOpen
  \bibfield  {author} {\bibinfo {author} {\bibfnamefont {C.~E.}\ \bibnamefont
  {Zaspel}}, \bibinfo {author} {\bibfnamefont {E.~S.}\ \bibnamefont {Wright}},
  \bibinfo {author} {\bibfnamefont {A.~Y.}\ \bibnamefont {Galkin}}, \ and\
  \bibinfo {author} {\bibfnamefont {B.~A.}\ \bibnamefont {Ivanov}},\ }\href
  {\doibase 10.1103/PhysRevB.80.094415} {\bibfield  {journal} {\bibinfo
  {journal} {Phys. Rev. B}\ }\textbf {\bibinfo {volume} {80}},\ \bibinfo {eid}
  {094415} (\bibinfo {year} {2009})}\BibitemShut {NoStop}%
\bibitem [{\citenamefont {Galkin}\ and\ \citenamefont
  {Ivanov}(2009)}]{Galkin09}%
  \BibitemOpen
  \bibfield  {author} {\bibinfo {author} {\bibfnamefont {A.}~\bibnamefont
  {Galkin}}\ and\ \bibinfo {author} {\bibfnamefont {B.}~\bibnamefont
  {Ivanov}},\ }\href {http://dx.doi.org/10.1134/S1063776109070103} {\bibfield
  {journal} {\bibinfo  {journal} {JETP}\ }\textbf {\bibinfo {volume} {109}},\
  \bibinfo {pages} {74} (\bibinfo {year} {2009})}\BibitemShut {NoStop}%
\bibitem [{\citenamefont {Shuty\u{\i}}\ and\ \citenamefont
  {Sementsov}(2007)}]{Shutyi07}%
  \BibitemOpen
  \bibfield  {author} {\bibinfo {author} {\bibfnamefont {A.}~\bibnamefont
  {Shuty\u{\i}}}\ and\ \bibinfo {author} {\bibfnamefont {D.}~\bibnamefont
  {Sementsov}},\ }\href {http://dx.doi.org/10.1134/S106377610705010X}
  {\bibfield  {journal} {\bibinfo  {journal} {Journal of Experimental and
  Theoretical Physics}\ }\textbf {\bibinfo {volume} {104}},\ \bibinfo {pages}
  {758} (\bibinfo {year} {2007})},\ \bibinfo {note}
  {10.1134/S106377610705010X}\BibitemShut {NoStop}%
\bibitem [{\citenamefont {Lee}\ \emph {et~al.}(2004)\citenamefont {Lee},
  \citenamefont {Deac}, \citenamefont {Redon}, \citenamefont {Nozieres},\ and\
  \citenamefont {Dieny}}]{Lee04a}%
  \BibitemOpen
  \bibfield  {author} {\bibinfo {author} {\bibfnamefont {K.-J.}\ \bibnamefont
  {Lee}}, \bibinfo {author} {\bibfnamefont {A.}~\bibnamefont {Deac}}, \bibinfo
  {author} {\bibfnamefont {O.}~\bibnamefont {Redon}}, \bibinfo {author}
  {\bibfnamefont {J.-P.}\ \bibnamefont {Nozieres}}, \ and\ \bibinfo {author}
  {\bibfnamefont {B.}~\bibnamefont {Dieny}},\ }\href
  {http://dx.doi.org/10.1038/nmat1237} {\bibfield  {journal} {\bibinfo
  {journal} {Nat Mater}\ }\textbf {\bibinfo {volume} {3}},\ \bibinfo {pages}
  {877} (\bibinfo {year} {2004})}\BibitemShut {NoStop}%
\bibitem [{\citenamefont {Berkov}\ and\ \citenamefont {Gorn}(2005)}]{Berkov05}%
  \BibitemOpen
  \bibfield  {author} {\bibinfo {author} {\bibfnamefont {D.}~\bibnamefont
  {Berkov}}\ and\ \bibinfo {author} {\bibfnamefont {N.}~\bibnamefont {Gorn}},\
  }\href {\doibase 10.1103/PhysRevB.71.052403} {\bibfield  {journal} {\bibinfo
  {journal} {Phys. Rev. B}\ }\textbf {\bibinfo {volume} {71}},\ \bibinfo
  {pages} {052403} (\bibinfo {year} {2005})}\BibitemShut {NoStop}%
\bibitem [{\citenamefont {Yang}\ \emph {et~al.}(2007)\citenamefont {Yang},
  \citenamefont {Zhang},\ and\ \citenamefont {Li}}]{Yang07a}%
  \BibitemOpen
  \bibfield  {author} {\bibinfo {author} {\bibfnamefont {Z.}~\bibnamefont
  {Yang}}, \bibinfo {author} {\bibfnamefont {S.}~\bibnamefont {Zhang}}, \ and\
  \bibinfo {author} {\bibfnamefont {Y.~C.}\ \bibnamefont {Li}},\ }\href
  {\doibase 10.1103/PhysRevLett.99.134101} {\bibfield  {journal} {\bibinfo
  {journal} {Phys. Rev. Lett.}\ }\textbf {\bibinfo {volume} {99}},\ \bibinfo
  {pages} {134101} (\bibinfo {year} {2007})}\BibitemShut {NoStop}%
\bibitem [{\citenamefont {Petit-Watelot}\ \emph {et~al.}(2012)\citenamefont
  {Petit-Watelot}, \citenamefont {Kim}, \citenamefont {Ruotolo}, \citenamefont
  {Otxoa}, \citenamefont {Bouzehouane}, \citenamefont {Grollier}, \citenamefont
  {Vansteenkiste}, \citenamefont {Van~de Wiele}, \citenamefont {Cros},\ and\
  \citenamefont {Devolder}}]{Petit-Watelot12}%
  \BibitemOpen
  \bibfield  {author} {\bibinfo {author} {\bibfnamefont {S.}~\bibnamefont
  {Petit-Watelot}}, \bibinfo {author} {\bibfnamefont {J.-V.}\ \bibnamefont
  {Kim}}, \bibinfo {author} {\bibfnamefont {A.}~\bibnamefont {Ruotolo}},
  \bibinfo {author} {\bibfnamefont {R.~M.}\ \bibnamefont {Otxoa}}, \bibinfo
  {author} {\bibfnamefont {K.}~\bibnamefont {Bouzehouane}}, \bibinfo {author}
  {\bibfnamefont {J.}~\bibnamefont {Grollier}}, \bibinfo {author}
  {\bibfnamefont {A.}~\bibnamefont {Vansteenkiste}}, \bibinfo {author}
  {\bibfnamefont {B.}~\bibnamefont {Van~de Wiele}}, \bibinfo {author}
  {\bibfnamefont {V.}~\bibnamefont {Cros}}, \ and\ \bibinfo {author}
  {\bibfnamefont {T.}~\bibnamefont {Devolder}},\ }\href
  {http://dx.doi.org/10.1038/nphys2362} {\bibfield  {journal} {\bibinfo
  {journal} {Nat Phys}\ }\textbf {\bibinfo {volume} {8}},\ \bibinfo {pages}
  {682} (\bibinfo {year} {2012})}\BibitemShut {NoStop}%
\bibitem [{\citenamefont {Kocarev}\ and\ \citenamefont
  {Parlitz}(1995)}]{Kocarev95}%
  \BibitemOpen
  \bibfield  {author} {\bibinfo {author} {\bibfnamefont {L.}~\bibnamefont
  {Kocarev}}\ and\ \bibinfo {author} {\bibfnamefont {U.}~\bibnamefont
  {Parlitz}},\ }\href {\doibase 10.1103/PhysRevLett.74.5028} {\bibfield
  {journal} {\bibinfo  {journal} {Phys. Rev. Lett.}\ }\textbf {\bibinfo
  {volume} {74}},\ \bibinfo {pages} {5028} (\bibinfo {year}
  {1995})}\BibitemShut {NoStop}%
\bibitem [{\citenamefont {\and Heinz Georg~Schuster}(2008)}]{Schoell08}%
  \BibitemOpen
  \bibinfo {editor} {\bibfnamefont {E.~S.}\ \bibnamefont {\and Heinz
  Georg~Schuster}},\ ed.,\ \href@noop {} {\emph {\bibinfo {title} {Handbook of
  Chaos Control}}},\ \bibinfo {edition} {2nd}\ ed.\ (\bibinfo  {publisher}
  {John Wiley \& Sons},\ \bibinfo {year} {2008})\ p.\ \bibinfo {pages}
  {849}\BibitemShut {NoStop}%
\bibitem [{\citenamefont {Ishida}\ \emph {et~al.}(2006)\citenamefont {Ishida},
  \citenamefont {Kimura},\ and\ \citenamefont {Otani}}]{Ishida06}%
  \BibitemOpen
  \bibfield  {author} {\bibinfo {author} {\bibfnamefont {T.}~\bibnamefont
  {Ishida}}, \bibinfo {author} {\bibfnamefont {T.}~\bibnamefont {Kimura}}, \
  and\ \bibinfo {author} {\bibfnamefont {Y.}~\bibnamefont {Otani}},\ }\href
  {http://link.aps.org/abstract/PRB/v74/e014424} {\bibfield  {journal}
  {\bibinfo  {journal} {Phys. Rev. B}\ }\textbf {\bibinfo {volume} {74}},\
  \bibinfo {eid} {014424} (\bibinfo {year} {2006})}\BibitemShut {NoStop}%
\bibitem [{\citenamefont {Muxworthy}\ \emph {et~al.}(2003)\citenamefont
  {Muxworthy}, \citenamefont {Dunlop},\ and\ \citenamefont
  {Williams}}]{Muxworthy03}%
  \BibitemOpen
  \bibfield  {author} {\bibinfo {author} {\bibfnamefont {A.~R.}\ \bibnamefont
  {Muxworthy}}, \bibinfo {author} {\bibfnamefont {D.~J.}\ \bibnamefont
  {Dunlop}}, \ and\ \bibinfo {author} {\bibfnamefont {W.}~\bibnamefont
  {Williams}},\ }\href {http://dx.doi.org/10.1029/2002JB002195} {\bibfield
  {journal} {\bibinfo  {journal} {J. Geophys. Res.}\ }\textbf {\bibinfo
  {volume} {108}},\ \bibinfo {pages} {2281} (\bibinfo {year}
  {2003})}\BibitemShut {NoStop}%
\bibitem [{\citenamefont {Kamionka}\ \emph {et~al.}(2011)\citenamefont
  {Kamionka}, \citenamefont {Martens}, \citenamefont {Drews}, \citenamefont
  {Kr\"uger}, \citenamefont {Albrecht},\ and\ \citenamefont
  {Meier}}]{Kamionka11}%
  \BibitemOpen
  \bibfield  {author} {\bibinfo {author} {\bibfnamefont {T.}~\bibnamefont
  {Kamionka}}, \bibinfo {author} {\bibfnamefont {M.}~\bibnamefont {Martens}},
  \bibinfo {author} {\bibfnamefont {A.}~\bibnamefont {Drews}}, \bibinfo
  {author} {\bibfnamefont {B.}~\bibnamefont {Kr\"uger}}, \bibinfo {author}
  {\bibfnamefont {O.}~\bibnamefont {Albrecht}}, \ and\ \bibinfo {author}
  {\bibfnamefont {G.}~\bibnamefont {Meier}},\ }\href {\doibase
  10.1103/PhysRevB.83.224424} {\bibfield  {journal} {\bibinfo  {journal} {Phys.
  Rev. B}\ }\textbf {\bibinfo {volume} {83}},\ \bibinfo {pages} {224424}
  (\bibinfo {year} {2011})}\BibitemShut {NoStop}%
\bibitem [{\citenamefont {Depondt}\ and\ \citenamefont
  {L\'{e}vy}(2012)}]{Depondt12}%
  \BibitemOpen
  \bibfield  {author} {\bibinfo {author} {\bibfnamefont {P.}~\bibnamefont
  {Depondt}}\ and\ \bibinfo {author} {\bibfnamefont {J.-C.}\ \bibnamefont
  {L\'{e}vy}},\ }\href {\doibase 10.1016/j.physleta.2012.09.039} {\bibfield
  {journal} {\bibinfo  {journal} {Physics Letters A}\ }\textbf {\bibinfo
  {volume} {376}},\ \bibinfo {pages} {3411 } (\bibinfo {year}
  {2012})}\BibitemShut {NoStop}%
\bibitem [{\citenamefont {Machado}\ \emph {et~al.}(2012)\citenamefont
  {Machado}, \citenamefont {Rappoport},\ and\ \citenamefont
  {Sampaio}}]{Machado12}%
  \BibitemOpen
  \bibfield  {author} {\bibinfo {author} {\bibfnamefont {T.~S.}\ \bibnamefont
  {Machado}}, \bibinfo {author} {\bibfnamefont {T.~G.}\ \bibnamefont
  {Rappoport}}, \ and\ \bibinfo {author} {\bibfnamefont {L.~C.}\ \bibnamefont
  {Sampaio}},\ }\href {\doibase 10.1063/1.3694757} {\bibfield  {journal}
  {\bibinfo  {journal} {Applied Physics Letters}\ }\textbf {\bibinfo {volume}
  {100}},\ \bibinfo {eid} {112404} (\bibinfo {year} {2012})}\BibitemShut
  {NoStop}%
\bibitem [{\citenamefont {Mihajlovic}\ \emph {et~al.}(2010)\citenamefont
  {Mihajlovic}, \citenamefont {Patrick}, \citenamefont {Pearson}, \citenamefont
  {Novosad}, \citenamefont {Bader}, \citenamefont {Field}, \citenamefont
  {Sullivan},\ and\ \citenamefont {Hoffmann}}]{Mihajlovic10}%
  \BibitemOpen
  \bibfield  {author} {\bibinfo {author} {\bibfnamefont {G.}~\bibnamefont
  {Mihajlovic}}, \bibinfo {author} {\bibfnamefont {M.~S.}\ \bibnamefont
  {Patrick}}, \bibinfo {author} {\bibfnamefont {J.~E.}\ \bibnamefont
  {Pearson}}, \bibinfo {author} {\bibfnamefont {V.}~\bibnamefont {Novosad}},
  \bibinfo {author} {\bibfnamefont {S.~D.}\ \bibnamefont {Bader}}, \bibinfo
  {author} {\bibfnamefont {M.}~\bibnamefont {Field}}, \bibinfo {author}
  {\bibfnamefont {G.~J.}\ \bibnamefont {Sullivan}}, \ and\ \bibinfo {author}
  {\bibfnamefont {A.}~\bibnamefont {Hoffmann}},\ }\href {\doibase
  10.1063/1.3360841} {\bibfield  {journal} {\bibinfo  {journal} {Appl. Phys.
  Lett.}\ }\textbf {\bibinfo {volume} {96}},\ \bibinfo {eid} {112501} (\bibinfo
  {year} {2010})}\BibitemShut {NoStop}%
\bibitem [{uni()}]{unicc}%
  \BibitemOpen
  \href {http://cluster.univ.kiev.ua/eng/} {}\bibinfo {note} {Kyiv {N}ational
  {T}aras {S}hevchenko {U}niversity high--performance computing cluster,
  \url{http://cluster.univ.kiev.ua/eng/}}\BibitemShut {NoStop}%
\bibitem [{btr()}]{btrzx}%
  \BibitemOpen
  \href {http://www.rz.uni-bayreuth.de/} {}\bibinfo {note} {Bayreuth
  {U}niversity computing cluster,
  \url{http://www.rz.uni-bayreuth.de/}}\BibitemShut {NoStop}%
\bibitem [{\citenamefont {Chen}(2004)}]{Chen04}%
  \BibitemOpen
  \bibfield  {author} {\bibinfo {author} {\bibfnamefont {W.-K.}\ \bibnamefont
  {Chen}},\ }\href@noop {} {\emph {\bibinfo {title} {The Electrical Engineering
  Handbook}}},\ edited by\ \bibinfo {editor} {\bibfnamefont {W.-K.}\
  \bibnamefont {Chen}}\ (\bibinfo  {publisher} {Elsevier Academic Press},\
  \bibinfo {year} {2004})\BibitemShut {NoStop}%
\end{thebibliography}
%
%
%

%


\end{document}